\documentclass{aa}
\usepackage{txfonts,textcomp}
\usepackage{graphicx}
\usepackage{color}
\usepackage{tikz}
\usepackage{amssymb}
\usetikzlibrary{shapes,arrows}
\usepackage{multirow}
\usepackage{natbib}
\usepackage{pdflscape}
\bibpunct{(}{)}{;}{a}{}{,}

\begin{document} 

\title{{\it Kepler\/} sheds new and unprecedented light on the variability of a
  blue supergiant: gravity waves in the O9.5Iab star HD\,188209\thanks{Based on
    photometric observations made with the NASA {\it Kepler\/} satellite and on
    spectroscopic observations made with four telescopes: the Nordic Optical
    Telescope operated by NOTSA and the Mercator Telescope operated by the
    Flemish Community, both at the Observatorio del Roque de los Muchachos (La
    Palma, Spain) of the Instituto de Astrof\'{\i}sica de Canarias, the T13 2.0m
    Automatic Spectroscopic Telescope (AST) operated by Tennessee State
    University at the Fairborn Observatory, and the Hertzsprung SONG telescope
    operated on the Spanish Observatorio del Teide on the island of Tenerife by
    the Aarhus and Copenhagen Universities and by the Instituto de Astrof\'isica
    de Canarias, Spain}}

\author{C. Aerts\inst{1,2} \and 
S.\ S\'{\i}mon-D\'{\i}az\inst{3,4} \and
S.\ Bloemen\inst{1,2} \and
J.\ Debosscher\inst{1}\and
P.\ I.\  P\'apics\inst{1}  \and 
S.\ Bryson\inst{5}\and
M.\ Still\inst{5,6}\and
E.\  Moravveji\inst{1}\and
M.\ H.\ Williamson\inst{7}\and
F.\ Grundahl\inst{8}\and
M.\ Fredslund Andersen\inst{8}\and
V.\ Antoci\inst{8}\and
P.~L.\ Pall\'e\inst{3,4}\and
J.\ Christensen-Dalsgaard\inst{8}\and
T.~M.\ Rogers\inst{9,10}
}

\institute{Instituut voor Sterrenkunde, KU Leuven, Celestijnenlaan 200D, 3001
  Leuven, Belgium\\ \email{Conny.Aerts@ster.kuleuven.be}
           \and
Department of Astrophysics/IMAPP, Radboud University Nijmegen,
             6500 GL Nijmegen, The Netherlands
           \and
Instituto de Astrof\'{\i}sica de Canarias, 38200, La Laguna, Tenerife, Spain
           \and
Departamento de Astrof\'{\i}sica, Universidad de La Laguna, 38205, La Laguna,
Tenerife, Spain \and
NASA Ames Research Center, Moffett Field, CA 94095, USA
\and
Bay Area Environmental Research Institute, 560 Third Street W., Sonoma, CA 95476, USA
\and
Center of Excellence in Information Systems, Tennessee State University, 3500
John A. Merritt Blvd., Box 9501, Nashville, TN 37209, USA
\and
Stellar Astrophysics Centre, Department of Physics and Astronomy, Aarhus
University, DK-8000 Aarhus C, Denmark
\and
Department of Mathematics and Statistics, Newcastle University, UK
\and
Planetary Science Institute, Tucson, AZ 85721, USA
}

\date{Received ; Accepted}
   
\titlerunning{{\it Kepler\/} photometry and high-resolution spectroscopy
of the O9.5 supergiant HD\,188209}
\authorrunning{C.\ Aerts et al.}

% \abstract{}{}{}{}{} 
% 5 {} token are mandatory
 
\abstract{Stellar evolution models are most uncertain for evolved massive
  stars. Asteroseismology based on high-precision uninterrupted space photometry
  has become a new way to test the outcome of stellar evolution theory and was
  recently applied to a multitude of stars, but not yet to massive evolved
  supergiants.Our aim is to detect, analyse and interpret the photospheric
  and wind variability of the O9.5\,Iab star HD\,188209 from {\it Kepler\/}
  space photometry and long-term high-resolution spectroscopy. We used {\it
    Kepler\/} scattered-light photometry obtained by the nominal mission during
  1460\,d to deduce the photometric variability of this O-type supergiant. In
  addition, we assembled and analysed high-resolution high signal-to-noise
  spectroscopy taken with four spectrographs during some 1800\,d to interpret
  the temporal spectroscopic variability of the star. The variability of
  this blue supergiant derived from the scattered-light space photometry is in
  full in agreement with the one found in the ground-based spectroscopy.  We
  find significant low-frequency variability that is consistently detected in
  all spectral lines of HD\,188209.  The photospheric variability propagates
  into the wind, where it has similar frequencies but slightly higher
  amplitudes. The morphology of the frequency spectra derived from the
  long-term photometry and spectroscopy points towards a spectrum of travelling
  waves with frequency values in the range expected for an evolved O-type
  star. Convectively-driven internal gravity waves excited in the stellar
  interior offer the most plausible explanation of the detected variability.}


\maketitle

%%%%%%%%%%%%%%%%%%%%%%%%%%%%%%%%%%%%%%%%%

\section{Introduction}

Stars born with sufficiently high mass to explode as supernova at the end of
their life have major impact on the dynamical and chemical evolution of
galaxies. Appropriate models of such pre-supernovae are thus highly relevant for
astrophysics.  Unfortunately, the theory of their evolution is a lot less well
established than the one of low-mass stars that die as white dwarf.  Differences
in the predictions of massive star evolution from various modern stellar
evolution codes even occur already well before the end of the main-sequence (MS)
phase \citep[e.g.,][]{Martins2013}.

Despite the immense progress in the asteroseismic tuning of stellar models of
various types of stars from high-precision uninterrupted space photometry in the
past decade \citep[e.g.,][for
reviews]{Chaplin2013,Charpinet2014,Aerts2015,Hekker2016}, we still lack suitable
data to achieve this stage for massive O-type stars and their evolved
descendants, the B supergiants.  Indeed, while the MOST and CoRoT missions did
observe a few B supergiants for weeks to months
\citep[e.g.,][]{Saio2006,Aerts2010,Moravveji2012,Aerts2013}, their pulsational
frequencies were not measured with sufficient precision and/or the angular
wavenumbers $(\ell,m)$ of their oscillation modes could not be identified
\citep[e.g.,][for a detailed description of mode identification in
asteroseismology]{Aerts2010}.  A similar situation occurs for the earlier
phases, given the absence of suitable highly sampled space photometric data with
sufficiently long time base for O stars, despite appreciable efforts \citep[see,
e.g.,][for an updated summary]{Buysschaert2015}.  Hence, improvement of the
input physics adopted in stellar models representing various evolutionary phases
of massive stars is still beyond reach, in contrast to such achievements for
low-mass stars \citep[e.g.,][to mention just a few
studies]{Bedding2011,Bischoff2011,Foster2015,Deheuvels2015}.

There are several reasons why asteroseismology of evolved stars in the mass
range of supernova-progenitors, i.e., with birth masses above some 8\,M$_\odot$,
is so difficult to achieve. First and foremost, such stars have large radii and
connected with this, their oscillation periods are several to tens of days. Any
multiperiodic nonradial mode beating pattern therefore reaches periods of years
and such a time base is beyond the capacity of the MOST, CoRoT, and K2 space
missions. This is also the reason why gravity-mode pulsators in the
core-hydrogen burning phase, which have periods of the order of half to a few
days, could only be fully exploited seismically thanks to the nominal {\it
  Kepler\/} mission.  Indeed, although their period-spacing pattern was first
discovered from 150\,d of uninterrupted CoRoT data \citep{Degroote2010}, seismic
modelling of MS gravity-mode pulsators required at least a year of uninterrupted
space photometry.  Four years of {\it Kepler\/} data of B and F stars led to
interior structure properties that cannot be explained with standard models, in
terms of interior rotation and mixing
\citep[e.g.,][]{Kurtz2014,Saio2015,Moravveji2015,Triana2015,Moravveji2016,Murphy2016,VanReeth2016,Schmid2016}. It
is thus to be expected that models of their evolved counterparts deviate even
more from reality.

A second reason that hampers asteroseismology of O stars and B supergiants is
the fact that their variability is not only caused by heat-driven coherent
stellar oscillations, but as well by a time-variable radiation-driven stellar
wind, rotational effects, macroturbulence, and for very few of the youngest O
stars magnetic activity as well \citep{Fossati2016}.  All these physical
phenomena interact, often non-linearly and in a non-adiabatic regime. This leads
to complex overall variability with even longer time bases than a
classical multiperiodic oscillation.

Here, we present a study of HD\,188209 (O9\,Iab), the only massive supergiant
that was monitored with the nominal {\it Kepler\/} mission during a total time
base of four years and about equally long in ground-based spectroscopy. We first
introduce the known properties of our target star and then discuss the long-term
monitoring in space photometry and in ground-based spectroscopy. We provide
evidence for variability with an entire spectrum of speriods of the order of half
to a few days in the independent data sets.

\begin{table*}[h!]
  \caption{Spectroscopically derived properties of HD\,188209 reported in the
    recent literature, based on a non-LTE analysis. Typical error bars are
    1\,000\,K for $T_{\rm eff}$, 0.2\,dex for $\log g$, and several km\,s$^{-1}$ for
    $v\sin i$ and 
    $v_{\rm macro}$.}
\label{pars}
\centering 
\tabcolsep=5pt                                    
\begin{tabular}{cccccccc}          
\hline\\[-10pt]     
Reference &  $T_{\rm eff}$ & $\log g$ &  [N(He)/N(H)] & 
$\log\dot{M}$&  $v_\infty$& $v\sin i$ &
$v_{\rm macro}$\\
 & (K) & (cgs) && (M$_\odot$\,yr$^{-1}$) & (km\,s$^{-1}$) & (km\,s$^{-1}$) & 
(km\,s$^{-1}$) \\
\hline\\[-10pt]
%\citet{Fullerton1996} & 30\,000& 3.3 & -6.10 & 1650 & 70 & --\\ 
%\citet{Israelian2000} & 31\,500 &3.0 & -5.80 & -- & -- & -- \\
\citet{Markova2005} & 31\,000 & 3.1 & 0.12 & -5.78 & 1650 & 87 & -- \\
%\citet{Fullerton2006} & 31\,000 & -- & -5.80 & 1650 & 92 & -- \\
%\cite{SimonDiaz2014} & -- & -- & -- & -- & 57 & 75 \\
\citet{Martins2015b,Martins2015a} & 29\,800 & 3.2 & -- & -6.40 & 2000 & 45 & 33 \\
Holgado et al.\ (in preparation) & 31\,100 & 3.0 & 0.14 & -- & -- &  57 & 75 \\
\end{tabular}
\end{table*}

\section{The O9\,Iab supergiant HD\,188209}

Given its visual magnitude of 5.63, HD\,188209 was the subject of various
observational variability studies so far. These mainly focused on spectroscopy
and were limited to only few spectra gathered with low sampling rates.  Early
spectroscopic time-series assembled by \citet{Fullerton1996} and
\citet{Israelian2000} revealed line-profile variability in the UV and optical
parts of its spectrum, with various time scales of the order of days. Additional
studies by \citet{Markova2005}, \citet{Fullerton2006}, and \citet{Martins2015b},
revealed variability in both the photosphere and in the stellar wind, with
seemingly uncorrelated quasi-periodicities occurring in those two
regimes.  

The spectral type assigned to HD\,188209 is O9.5Iab
\citep{Walborn1972,Sota2011}. It is included in the list of standard stars for
spectral classification \citep[e.g.,][]{Walborn1990, Maiz2015}.  The fundamental
parameters, mass-loss rate, and level of macroturbulence from recent analyses
based on the non-LTE codes 
CMFGEN \citep{Martins2015b,Martins2015a} and FASTWIND
\citep[][Holgado et al.\ 2017, in preparation]{Markova2005}, both including the
effects of line blanketing and the stellar wind, are listed in
Table\,\ref{pars}.  While binarity dominates the evolution of massive stars
\citep{Sana2012}, HD\,188209 was found to be a single star
\citep{Martins2015a}. In line with the low incidence of surface magnetic fields
in OB stars, HD\,188209 led to a non-detection of such a field
\citep{Grunhut2017}.

Given its presence in the nominal Field-of-View (FoV) of
the {\it Kepler\/} satellite, we have embarked on a unique long-term monitoring study
of HD\,188209, based on space photometry and ground-based high-resolution
spectroscopy. 

\section{Scattered-light {\it Kepler\/} photometry}

\begin{figure*}
\begin{center}
\rotatebox{0}{\resizebox{15cm}{!}{\includegraphics{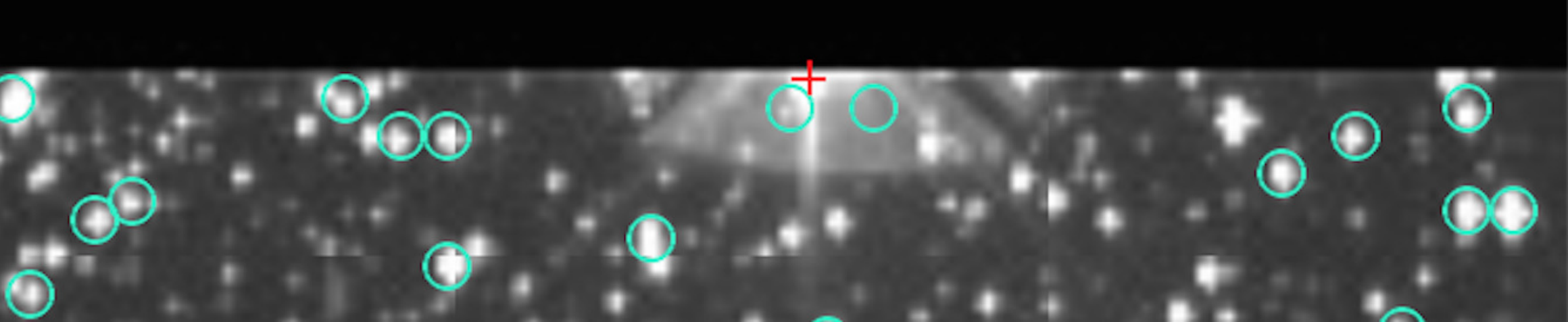}}}
\end{center}
\caption{Position of the two apertures (45: left), 32 (right) in the {\it Kepler\/} FoV 
capturing the scattered light of HD\,188209 (situated in the black band 
above the red cross in
between active silicon).}
\label{CCD}
\end{figure*}

\begin{figure*}
\begin{center}
\rotatebox{0}{\resizebox{9cm}{!}{\includegraphics{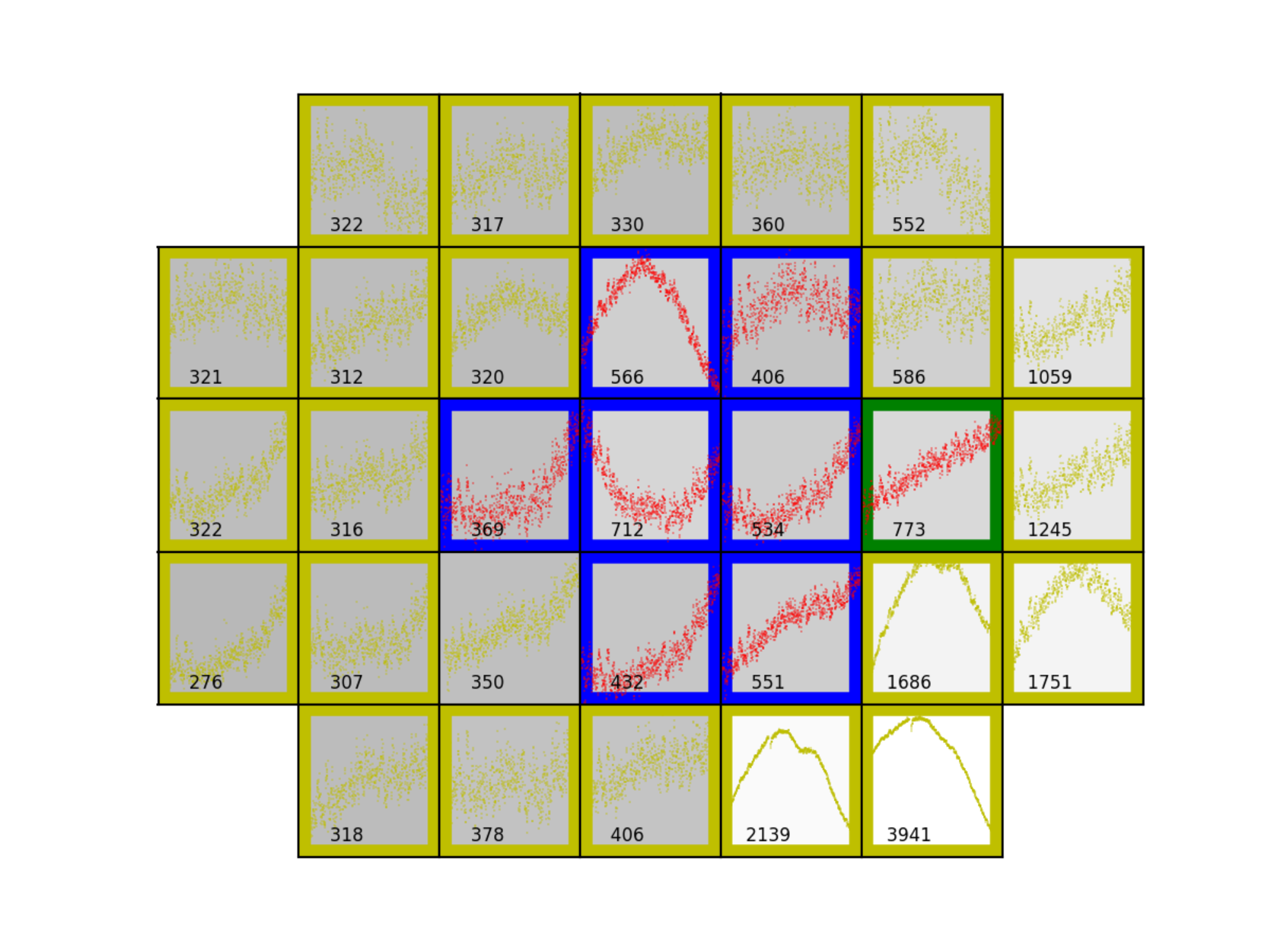}}}
\rotatebox{0}{\resizebox{9cm}{!}{\includegraphics{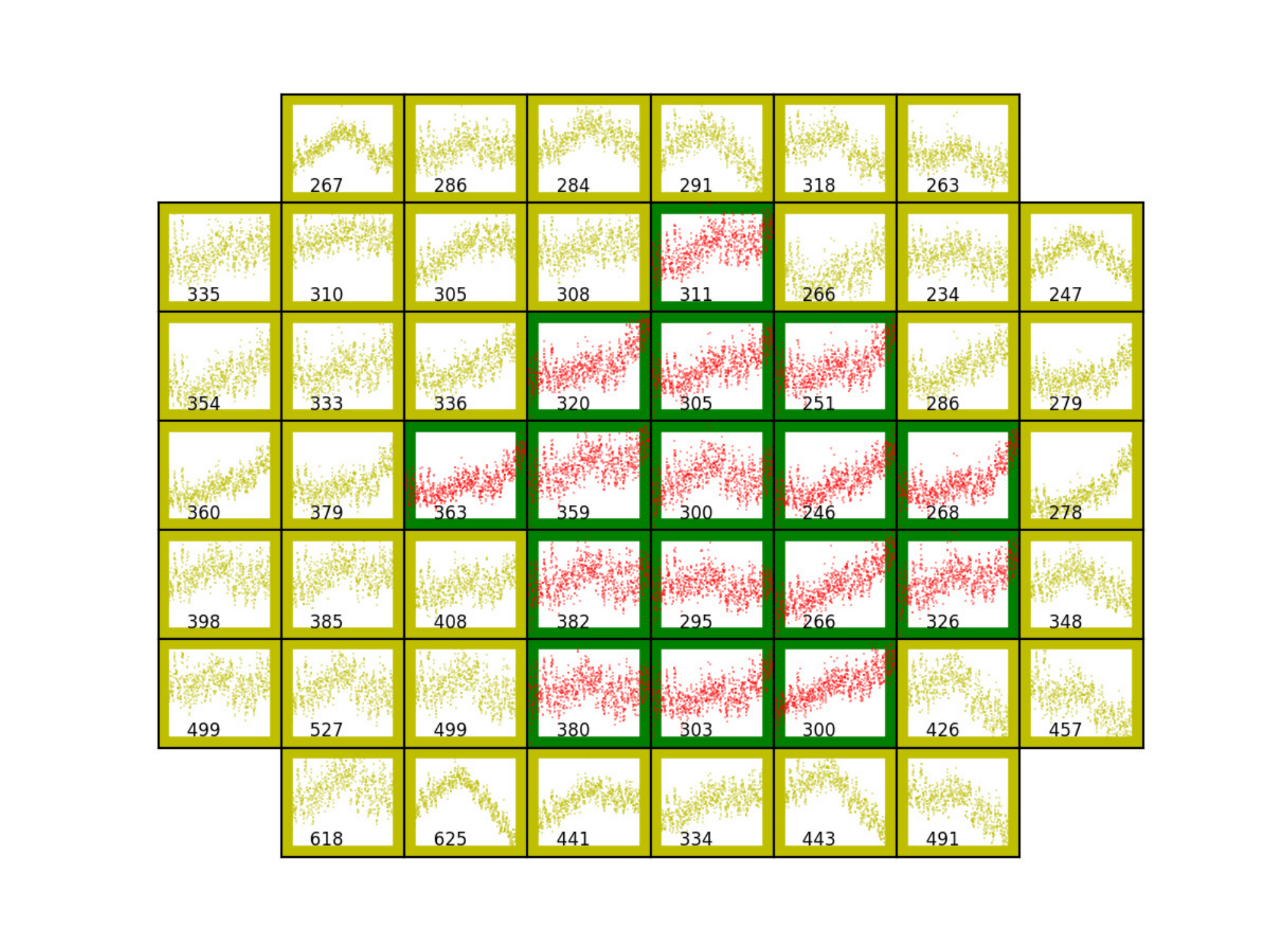}}}
\end{center}
\caption{An example of pixel masks used to create three versions of the
    light curves for star 45 (left) and 32 (right).  The detected light curve is
    plotted within each pixel and represents an addition of 270 individual
    exposures of 6.54\,s, added onboard of the satellite. In long-cadence mode,
    such pixel data are downloaded from the spacecraft in time stamps of 29.42
    minutes.  The value listed in every pixel, as well as the background grey
    scale, indicate the SNR of the flux in the pixel.  The red dots in the
    pixels with green and blue borders were used to extract the standard {\it
      Kepler\/} light curves. We also computed light curves based on the data in
    the yellow pixels alone and from adding the green, blue, and yellow pixels.}
\label{masks}
\end{figure*}

\begin{figure}
\begin{center}
\rotatebox{0}{\resizebox{9.cm}{!}{\includegraphics{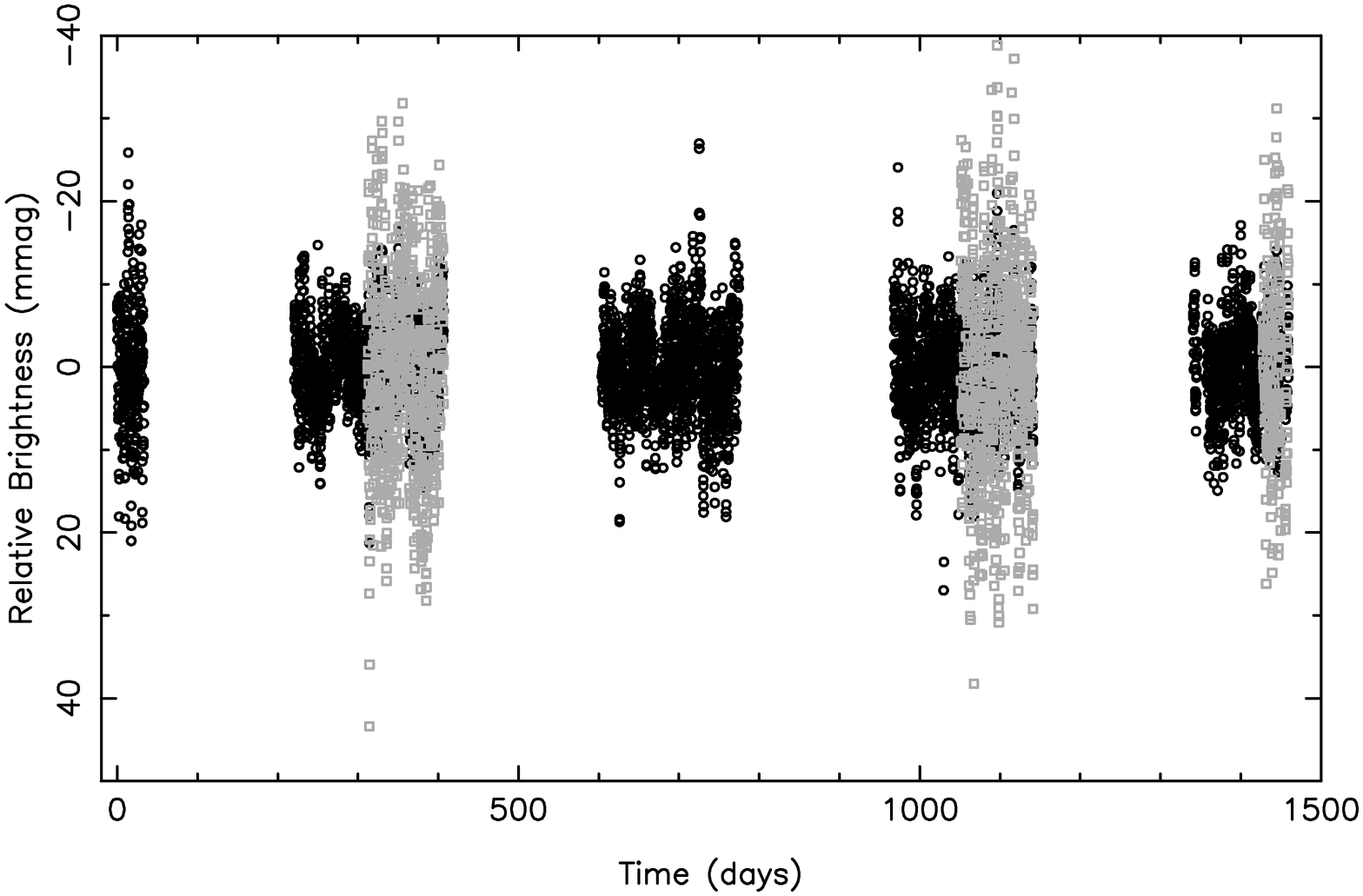}}}
%\vspace{0.3cm}
\rotatebox{0}{\resizebox{9.cm}{!}{\includegraphics{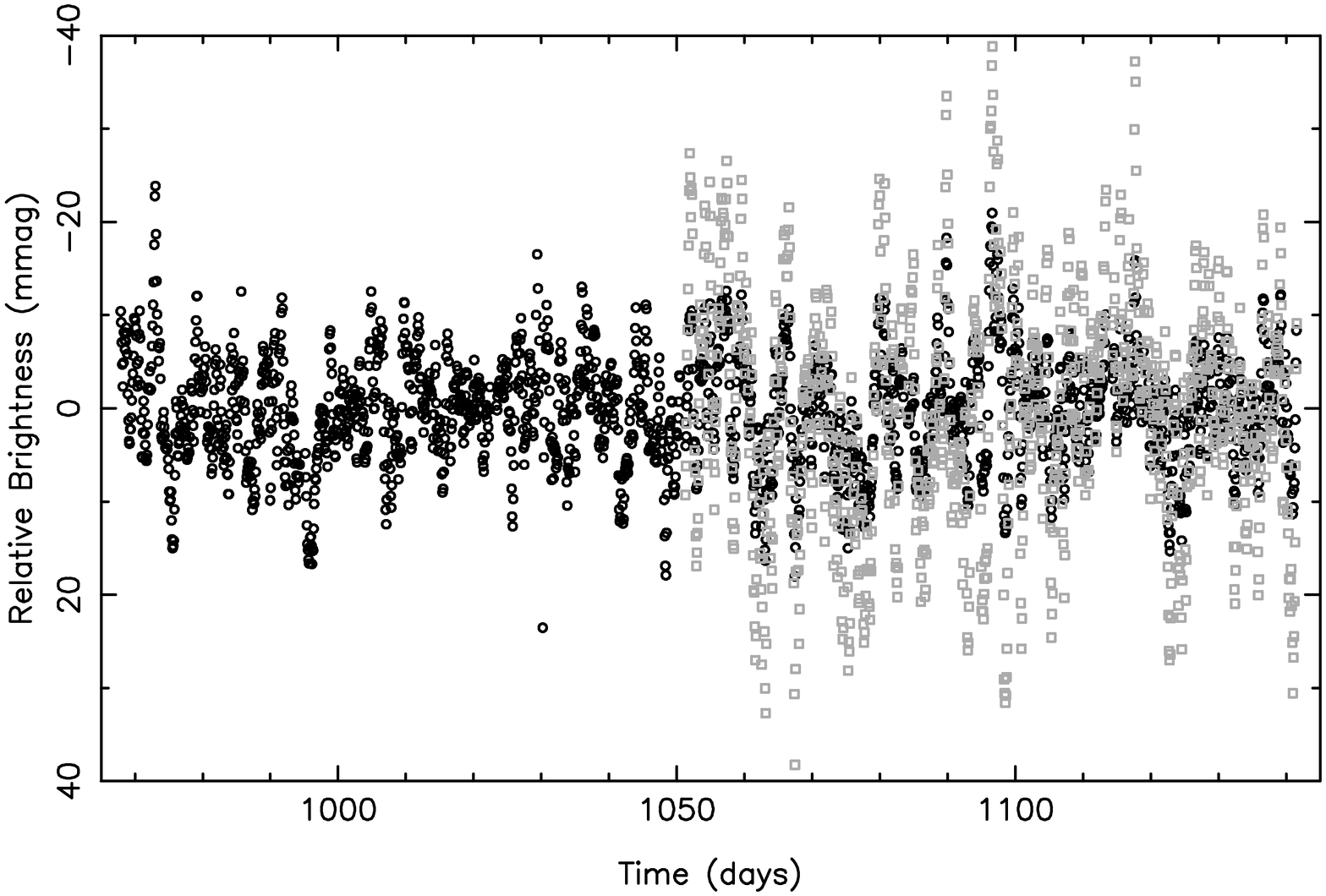}}}
%\vspace{0.3cm}
\rotatebox{270}{\resizebox{5.6cm}{!}{\includegraphics{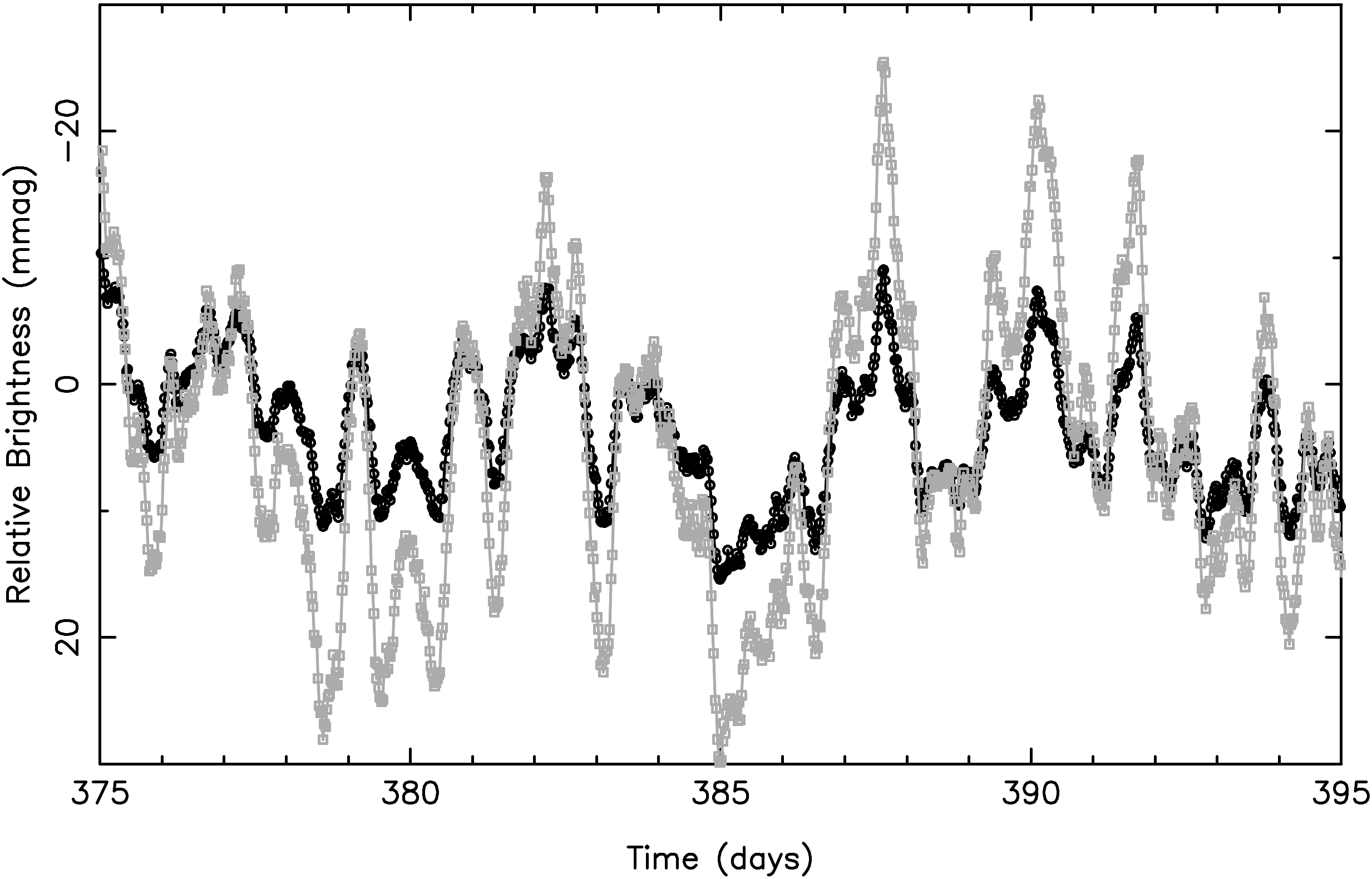}}}
\end{center}
\caption{Light curves in masks 32 (grey circles) and 45 (black circles) covering
  the full length (upper), quarters 4 and 5 (middle) and a zoom of 20\,d.}
\label{LCs}
\end{figure}

\begin{figure}
\begin{center}
\rotatebox{0}{\resizebox{9.cm}{!}{\includegraphics{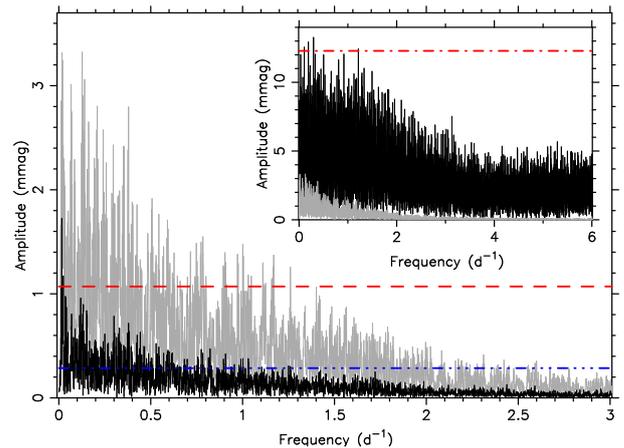}}}
\end{center}
\caption{Amplitude spectrum for the {\it Kepler\/} light curves of 
masks 32 (grey) and 45 (black). The inset compares the spectrum
of the {\it Kepler\/} light curve from  
mask 32 (grey) with the {\it Hipparcos\/} light curve (black) over twice the
frequency range. The red dashed and blue dot-dot-dot-dashed lines indicate
four times the average noise level computed over the frequency range 
$[0,6]\,$d$^{-1}$, while the red
dot-dashed line in the inset represents  
four times the average noise level computed over the same range 
$[0,6]\,$d$^{-1}$ of the {\it Hipparcos\/} data.}
\label{Scargle-Kp-Hp}
\end{figure}

\begin{figure}
\begin{center}
\rotatebox{0}{\resizebox{9cm}{!}{\includegraphics{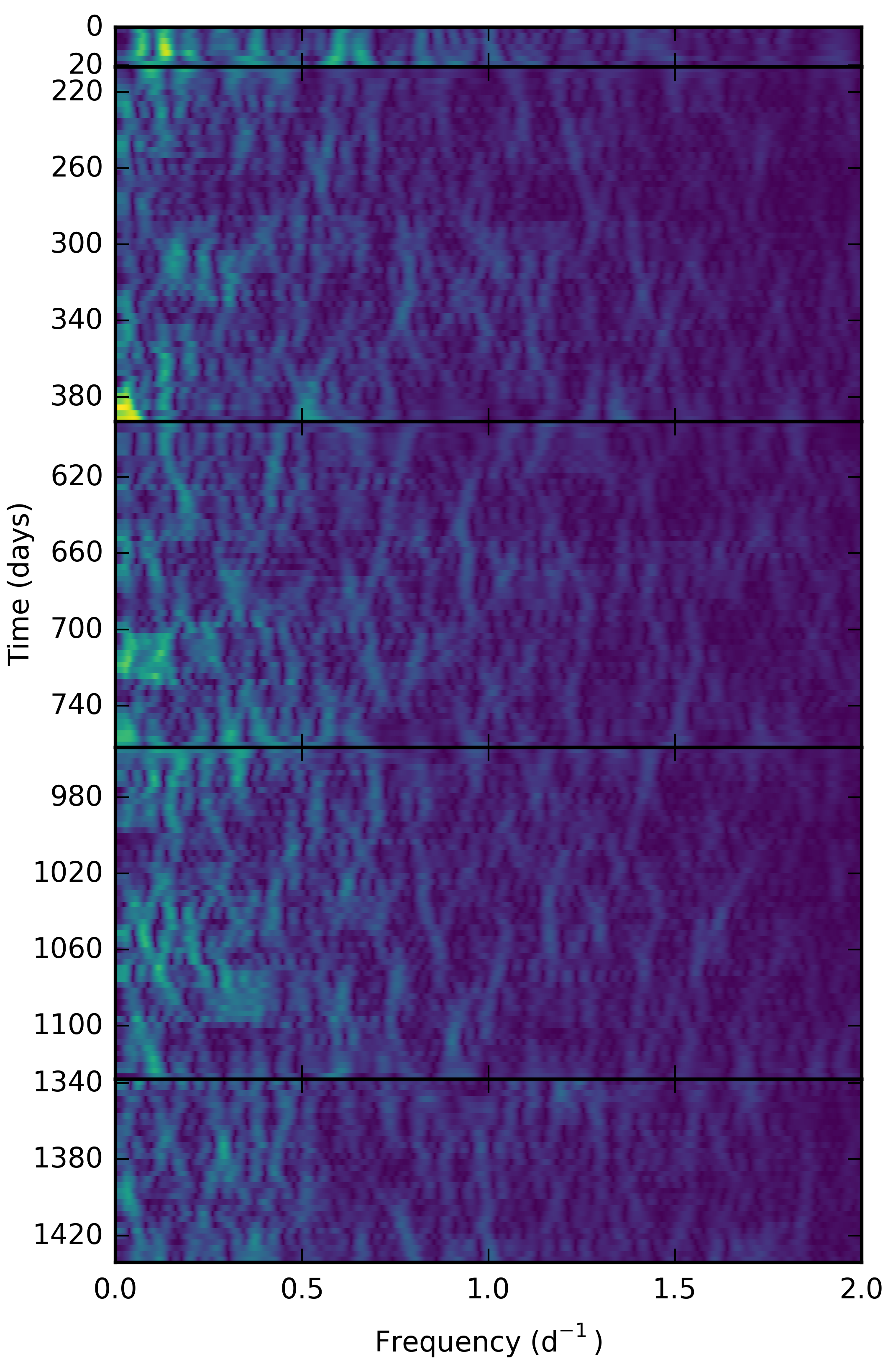}}}\hspace{0.3cm}
\rotatebox{0}{\resizebox{9cm}{!}{\includegraphics{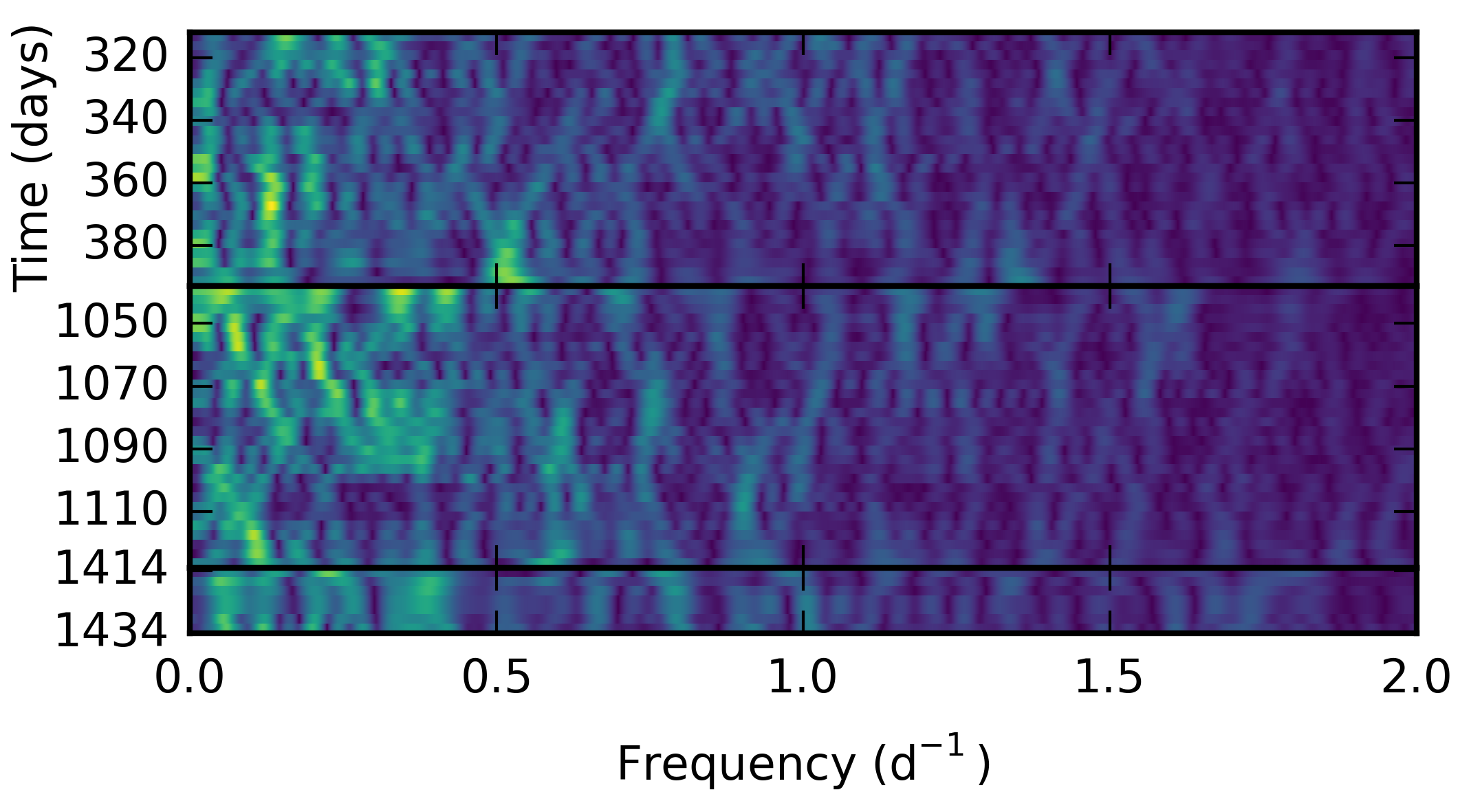}}}
\end{center}
\caption{Short-time Fourier transforms of the {\it Kepler\/} light curves in
  mask 45 (upper) and 32 (lower) for all observed quarters, for a 30-day time
  window and progressing in the time series with a step of 3 days, zoomed into
  the region of low-frequency amplitude excess. Brighter colours represent
  higher amplitudes.}
\label{stft}
\end{figure}

The {\it Kepler\/} CCDs saturate for $K_p\sim\,11.5$, the precise value
depending on the CCD and the target location on that CCD
\citep{Koch2010}. However, it turned out that the saturated flux is conserved to
a very high degree as long as sufficiently large masks are placed around the
star under study. This led to several variability studies of bright stars with
unprecedented photometric precision and duration of the light curves by means of
customized masks
\citep[e.g.][]{Kolenberg2011,Metcalfe2012,Tkachenko2014,Guzik2016}.  Meanwile,
numerous bright stars are also studied with the refurbished {\it Kepler\/}
mission K2, from co-adding carefully masked smear flux in the CCD rows of data
taken with ultra-short exposure times (White et al., submitted), a technique
that was verified successfully on nominal {\it Kepler\/} data
\citep[e.g.,][]{Pope2016}.

Here, we present a different use of the original {\it Kepler\/} satellite,
following our idea to perform scattered-light aperture photometry of HD\,188209,
despite the star not being on active silicon.  With $K_p\sim\,5.5$,
HD\,188209 was indeed a nuissance for exoplanet hunting with the nominal {\it
  Kepler\/} mission and it was therefore placed carefully in between the active
CCDs pointing to the nominal FoV. Nevertheless, it is so bright
that its scattered light does ``pollute'' the CCDs (cf.\ Fig.\,\ref{CCD}). As
only targets on active silicon are eligible for the {\it Kepler\/} and K2 Guest
Observer programs, we had to proceed in a different way and requested 
``informally'' from the {\it Kepler\/} GO office
to place masks in the vicinity of HD\,188209, with the reasoning that its
variability behaviour will be present and detectable in the scattered-light
photometry.

For the current analysis, we used two apertures placed in the
scattered light of the star.  The two masks placed on active silicon correspond
to the targets labelled KIC\,10092932 (to the right) and KIC\,10092945 (to the
left) of HD\,188209 in Fig.\,\ref{CCD}, hereafter called mask 32 and mask 45,
respectively.  
Mask 45 contains a faint star of $K_p\sim\,13.35$, which is 7.85 magnitudes
  fainter than HD\,188209 itself; mask 32 does not contain any known source and
  is therefore a ``pure'' probe of the scattered light of the supergiant.  It
  has $K_p=11.2$ as ``brightness'' for the scattered light.  Both these
  masks were observed in long-cadence (LC) mode, with a total integration time of
  29.42 minutes per data point. Each downloaded LC flux value is an on-board
  addition of 270 individual exposures of 6.54\,s, with 0.52\,s readout
  time.

  Masks of a selected {\it Kepler\/} quarter showing the light curves in the
  individual pixels as downloaded from the spacecraft for each of the two
  apertures are shown in Fig.\,\ref{masks}.  These show the typical time series
  traces with pixel-to-pixel variations as for normal targetted stars
  \citep[e.g.,]{}{Papics2014}, except that the flux levels are lower here.  This
  is particularly so for mask 32 given that it concerns scattered light rather
  than an actual target.  The signal-to-noise ratio (SNR hereafter) of the
  detected flux in every pixel is indicated in each of the boxes.  The pixels
  with dark green or blue borders with the red dots as flux values (stored in
  e$^-$\,s$^{-1}$) were used to extract the ``standard'' {\it Kepler\/} light
  curves of targets as provided in MAST.\footnote{\tt
  archive.stsci.edu/kepler/data$_{-}$search/search.php} We did not only use
these light curves but also constructed {\bf two} new ones that are based on the
yellow pixels in which significant signal coming from the supergiant is present
as well; SNR values above 100 typically correspond to useful signal to be
  added for stellar variability studies. Adding the flux in those yellow pixels
  in addition to the standard MAST pixels leads to cleaner detrending and less
  contamination by instrumental effects at low frequencies
  \citep{Tkachenko2013,Papics2014}.  For each of the two custom masks, we
extracted two light curves in addition to the standard MAST curves: one
based on the yellow flux alone and another one based on the green+blue+yellow
flux. For each of the two masks, the variability behaviour of these 
  three light curves is the same in terms of periodic behaviour, which is
evidence that this variability is dominated by the scattered light of the
supergiant and not by other sources in the pixel masks.

In the rest of the paper, we proceed with the light curves containing maximal
signal-to-noise ratio (SNR hereafter). These are the ones deduced from adding
all the flux in the red, blue, and yellow pixels of 
masks 32 and 45 as shown in Fig.\,\ref{masks}.
Figure\,\ref{LCs} compares these two light curves, where mask 45 was observed
during 1460\,d from 13/5/2009 until 9/4/2013 (quarters 1,4,5,8,9,12,13,16,17; in
total 31,034 measurements) and mask 32 during 1147\,d from 20/3/2010 until
9/4/2013 (quarters 5,13,17; in total 10,080 measurements).  {\it Despite it
  being scattered light, this is by far the best light curve ever obtained for a
  blue supergiant.}  It can be seen in the lower panel of
Fig.\,\ref{LCs} that the two curves are very similar in terms of temporal
variability, but that the amplitude of the variability is different. This is as
expected given the difference in flux level within the masks and the fact that
mask 45 contains a faint star while mask 32 does not.

We computed the Fourier transform of the two light curves shown in
Fig.\,\ref{LCs} to obtain the amplitude spectra up to the Nyquist
frequency. This resulted in a flat distribution of noise beyond 6\,d$^{-1}$ and
a clear amplitude excess above the noise level at low frequency for both light
curves. A part of the spectrum is shown in Fig.\,\ref{Scargle-Kp-Hp}, where the
horizontal lines are situated at four times the average noise level
computed over the range $[0,6]\,$d$^{-1}$. Given that mask 32 is the ``purer''
one in terms of only scattered light of HD\,188209, its amplitude is dominant
over the one obtained from mask 45. However, both light curves are fully
consistent with each other in the sense that significant variability is detected
over the whole frequency range $[0,2]\,$d$^{-1}$ in the form of an excess of
amplitude for numerous close frequencies. This is a frequency regime 
typical for gravity waves in massive stars.

The amplitude spectrum in Fig.\,\ref{Scargle-Kp-Hp} is entirely different in
nature from the one of the rotationally variable B5 supergiant HD\,46769
observed with CoRoT during 23\,d and revealing only one frequency and its
harmonics corresponding to a rotation period of 4.8$\pm0.2$\,d
\citep{Aerts2013}. It also bears no resemblance to the frequency spectrum of the
B6 supergiant HD\,50064 whose 137-d CoRoT light curve revealed one dominant
radial pulsation mode with a period of 37\,d, identified as the fundamental
radial mode of the star \citep{Aerts2010}.  The morphology of the spectrum is
also different from those found for {\it Kepler\/} data of B dwarfs pulsating in
coherent heat-driven gravity modes \citep[e.g.,][]{Papics2015,Papics2016} as
well as from those of O dwarfs with variability assigned to rotational modulation or
heat-driven coherent non-radial pulsations \citep[see Table\,3 in][for an
overview]{Buysschaert2015}.

Since we are dealing with a photometric curve deduced from scattered light, a
comparison with other photometric data of the actual star is meaningful.  The
only publicly available light curve of HD\,188209 to make such a comparison is the one
assembled by the Hipparcos satellite, which is of similar total length but has
only 113 data points spread over 1156\,d, with far less precision than the {\it
  Kepler\/} scattered light data.  This data set was already analysed by
\citet[][see their Fig.\,3]{Israelian2000} and revealed variability but
did not lead to clear periodicity.  We recomputed the amplitude spectrum of the
Hipparcos light curve and compared it with the {\it Kepler\/} one of mask 32,
which has similar length, in the inset of Fig.\,\ref{Scargle-Kp-Hp}. Although
the significance of the amplitude excess in the Hipparcos data is only marginal
at best, low-frequency excess in $[0,2]\,$d$^{-1}$ is also hinted at in this
independent photometric data set, at a level of some 14\,mmag, while we found 
3.6\,mmag from the scattered light in mask 32.

The photometric amplitude spectrum of the scattered light of HD\,188209 is {\it
  remarkably similar\/} to the one obtained from the {\it Kepler\/} data of the
bright primary B0\,III star \citep{Lesh1968} of the eclipsing binary
V380\,Cyg. V380\,Cyg\,A's variability behaviour was interpreted in terms of
excess power due to internal gravity waves \citep[see Fig.\,3
in][]{Tkachenko2014}. This star has similar $\log\,g$ as HD\,188209 but is much
less massive (some 12\,M$_\odot$) and cooler (21,700\,K).  Similarly shaped
excess power, although covering a broader range in frequency, were also found
for three unevolved O stars hotter than HD\,188209 (effective temperatures
between 35,000\,K and 43,000\,K) observed by the CoRoT satellite
\citep{Blomme2011} and explained in terms of internal gravity waves caused by
core convection in \citet{AertsRogers2015}.  Even though further analysis of the
excitation layer of the waves -- convective core or convection zone in the
stellar envelope due to the iron opacity bump -- needs to be sought, variability
due to the stellar wind would results in few quasi-periodicities on a timescale
of a day to a week \citep{Kaper1996}, while here we are dealing with an {\it
  entire spectrum of excited low frequencies\/} without a dominant base frequency.  For
these four stars in the literature, the nature of the variability was
revealed through short-time Fourier transformations (STFTs; Fig.\,3 in
\citet{Tkachenko2012} and Fig.\,6 in \citet{Blomme2011} for these four stars).
Figure\,\ref{stft} shows STFTs of HD\,188209 based on the two {\it Kepler\/}
data sets. They were calculated using a 30-day time window, each time
progressing in the time series with a step of 3 days, but the results were
checked to remain the same for other values of the window and step.  Just as for
the four OB stars in the literature, it is clear from Fig.\,\ref{stft} that the
signal in these STFTs of HD\,188209 is not due to multimode beating of stable,
phase-coherent non-radial pulsation modes because the latter's STFTs look quite
different \citep[see, e.g., Fig.\,5 in][]{Degroote2012}.

We conclude that the {\it Kepler\/} scattered light photometry of HD\,188209
points towards a fifth case of highly significant variability with a
multitude of low frequencies for a hot massive star; it is the first case of a
massive supergiant for which such type of variability is revealed.  All five
stars in which this phenomenon has been found in high-cadence space photometry
are moderate rotators, having $v\sin i$ values between 50 and 100\,km\,s$^{-1}$
and require considerable macroturbulent broadening to explain the line-profile
shapes in high-resolution spectroscopy.  In order to properly interpret the
variability detected already early on in the scattered light of HD\,188209, we
initiated long-term ground-based follow-up high-resolution spectroscopy and
included the star in the IACOB project \citep{SimonDiaz2014}.

\section{Long-term high-resolution spectroscopy}

HD\,188209 was added to the large sample of OB-type stars across the entire
evolutionary path for long-term ground-based high signal-to-noise spectroscopic
monitoring \citep{SimonDiaz2015,SimonDiaz2017}. In contrast to the {\it
  Kepler\/} photometry, such type of data have a low duty cycle and limited time
coverage per night.  This inevitably leads to heavily gapped spectroscopic time
series, where the interpretation of the variability encounters the challenge to
deal with daily alias structures in the Fourier domain. Moreover, for a star as
HD\,188209, we will be dealing with quasi-periodicities occurring due to a
mixture of photospheric and wind variability. This was indeed already revealed
for HD\,188209 by \citet{Martins2015b,Martins2015a}, who studied it from a
time-series of spectro-polarimetry consisting of 27 spectra gathered over a time
span of 9\,d. They considered twelve spectral lines (three Balmer lines, seven
helium lines, and two metal lines) and found eight to be variable, including
those that form (partially) in the wind, while four photospheric lines turned
out to be stable in their data.  \citet{Martins2015b} concluded that the
photospheric line profiles of HD\,188209 seemingly change on time scales from an
hour to days, while its wind variability reveals longer-term periodicity
\citep[see also][]{Markova2005}.

No obvious connection between the quasi-periodicities in the photosphere and in
the wind was found for HD\,188209 so far. However, all the data sets available
in the literature have too few spectra, too sparse sampling, and too limited
time base to find periodicities with appropriate precision.  Given the limited
amount of spectra, all these studies only considered the variability inside the line
profile without performing frequency analysis, 
by means of the so-called temporal variance spectrum \citep{Fullerton1996}. This
quantity is very useful to estimate the overall level of variability that is
present in a small series of spectral lines assembled with sparse sampling. 
Its capacity to unravel the {\it cause\/}
of this variability is limited. Here, we shall be focusing on the {\it temporal
  line variability\/} in a large data set of spectra with a long time base,
with the aim to investigate the physical reasons of its
origin. We do this by considering specialised line-profile quantities that are
specifically designed to allow for time-series analysis without being affected
by uncertainties due to spectrum normalisation, as will be discussed below.

We take a major step ahead compared to the spectroscopy results in the 
literature from new extensive long-term spectroscopic
time series assembled with four instruments:
\begin{itemize}
\item the fiber-fed HERMES \'echelle spectrograph attached to the 1.2m Mercator
  telescope at Roque de los Muchachos, Island of La Palma, Spain, operated in
  the high-resolution (R=85,000) mode \citep{Raskin2011};
\item the high-resolution FIbre-fed \'Echelle Spectrograph (FIES) attached to
  the Nordic Optical Telescope (NOT) at Roque de los Muchachos, Island of La
  Palma, Spain, operated in the high-resolution (R=46,000) mode
  \citep{Telting2014};
\item the T13 2.0m Automatic Spectroscopic Telescope (AST) with the fiber-fed
  cross-dispersed \'echelle spectrograph in the resolution mode of R=30,000
  operated at the Fairborn Observatory, USA. A full description of the
  instrument and data reducation pipeline is available in \citet{Eaton2004};
\item the prototype SONG node \'echelle spectrograph, operational at Observatory
  del Teide on Tenerife, Spain
  \citep{Grundahl2007,Uytterhoeven2012,Grundahl2017}.  
The SONG
  observations of HD\,188209 were obtained in ThAr mode with slit 5 and have
  R=77\,000.
\end{itemize}
A summary of the characteristics of all the data sets analysed in this paper is
given in Table\,\ref{logbook}. 

\begin{table*}
\caption{Summary of the observations treated in this work; $N$ is the total
  number of measurements and $\Delta T$ is the total time base. Rayleigh is 
  equal to $1/(\Delta T)$ and Nyquist is
  set equal to $1/(2\Delta t)$, where $\Delta t$ is the minimal time
  between two consecutive exposures. }
\label{logbook}
\centering 
\tabcolsep=5pt                                    
\begin{tabular}{cccrrcc}
\hline\\[-10pt]     
Instrument &  HJD start & HJD end & $N$ & $\Delta T$ & Rayleigh & Nyquist \\
 & (-2450000) & (-2450000) &  & (d) & (d$^{-1}$) & (d$^{-1}$)\\
\hline\\[-10pt]
{\it Kepler\/}, Mask 32 & 5276.4899 & 6424.0009 & 10085 & 1147.5110 & 0.00087 & 
24.4738\\
{\it Kepler\/}, Mask 45 & 4964.5118 & 6424.0009 & 31074 & 1459.4891 & 0.00069 & 
24.4738\\
\hline\\[-10pt]
HERMES@Mercator & 5726.4523 & 7526.7411 & 102 & 1800.2887 & 0.00056 & 
26.5877\\ 
FIES@NOT & 5146.4053 & 5815.5758 & 11 & 669.1705 & 0.00149 & 8.2830\\ 
T13 2m AST & 6433.7684 & 6602.7555 & 228 & 168.9871 & 0.00592 & 46.9585\\
SONG & 7240.3758 & 7617.6022 & 417 & 377.2264 & 0.00265 & 47.8460 
\end{tabular}
\end{table*}

\begin{figure*}
\begin{center}
\rotatebox{270}{\resizebox{14.cm}{!}{\includegraphics{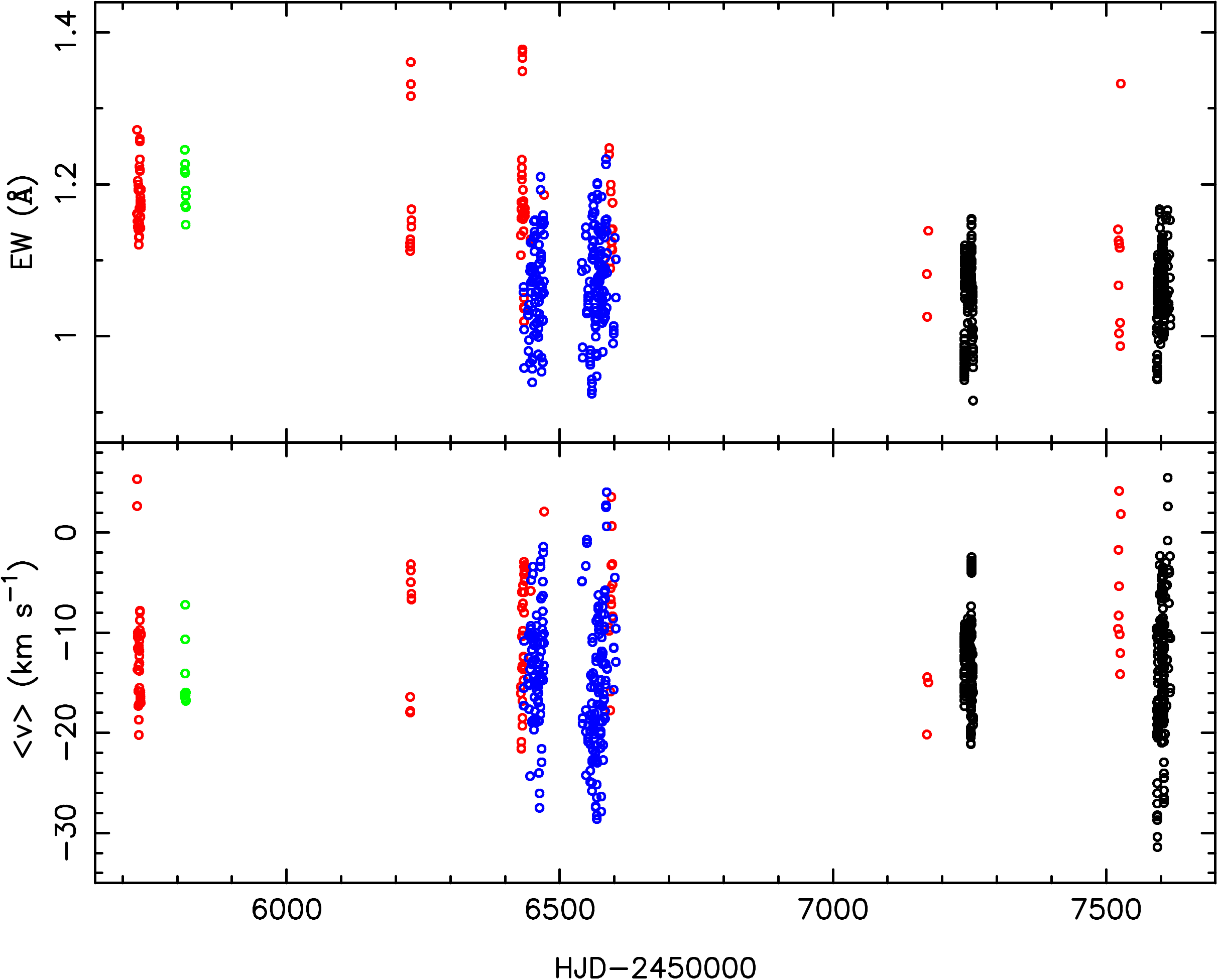}}}
\end{center}
\caption{The centroid velocity (lower panel) and equivalent width (upper panel) of
  the entire spectroscopic data set for the \ion{He}{i}\,5875\AA\ line,
  according to the following colour code: red=HERMES, 
green=FIES, blue=AST, and black=SONG.}
\label{vrad-ew-5875}
\end{figure*}

\begin{figure}
\begin{center}
\rotatebox{0}{\resizebox{9.cm}{!}{\includegraphics{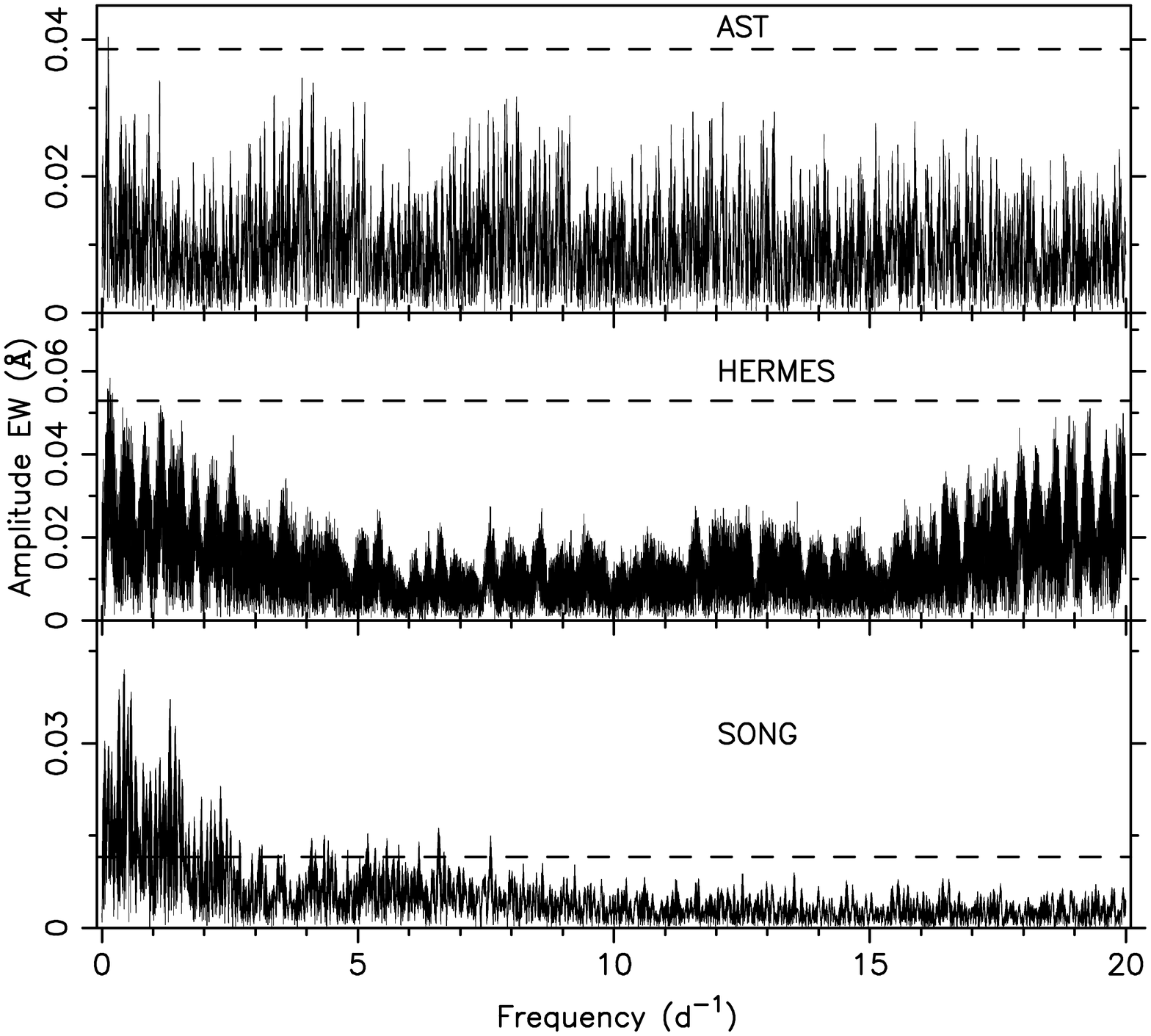}}}
%\vspace{0.3cm} 
\rotatebox{0}{\resizebox{11.cm}{!}{\includegraphics{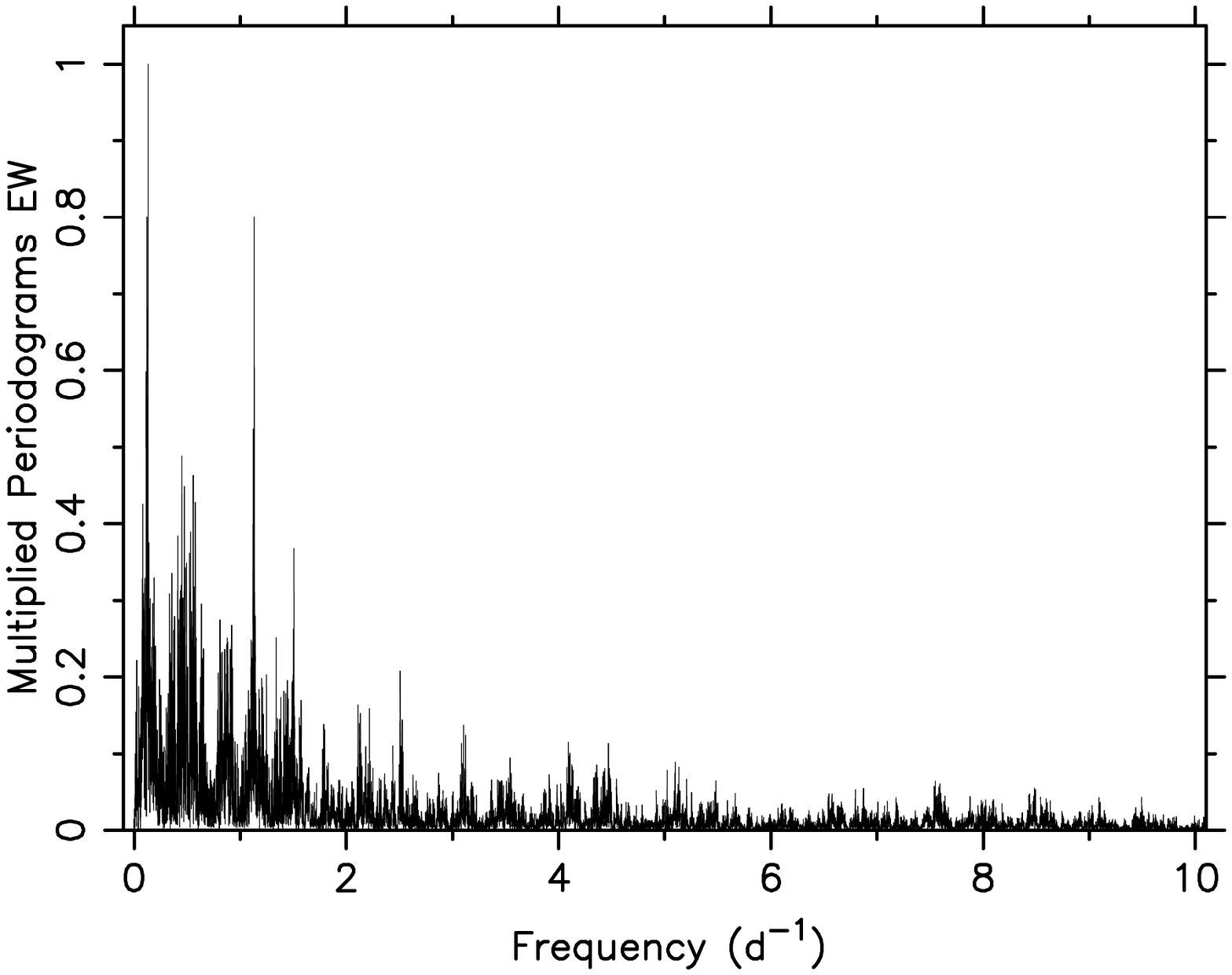}}}
\end{center}
\caption{Amplitude spectra for the EW of the \ion{He}{i}\,5875\AA\ line for the three
  individual spectrographs with sufficient data (upper panel, the red dashed
  lines indicate four times the average noise level computed over the range
  $[0,10]\,$d$^{-1}$). The lower panel shows the result of multiplying the
  periodograms in the upper panels after weighing them according to the SNR of
  the dominant frequency and placing the dominant frequency after this
  multiplication at value 1.}
\label{Scargle-ew-5875}
\end{figure}

\begin{figure}
\begin{center}
\rotatebox{0}{\resizebox{9.cm}{!}{\includegraphics{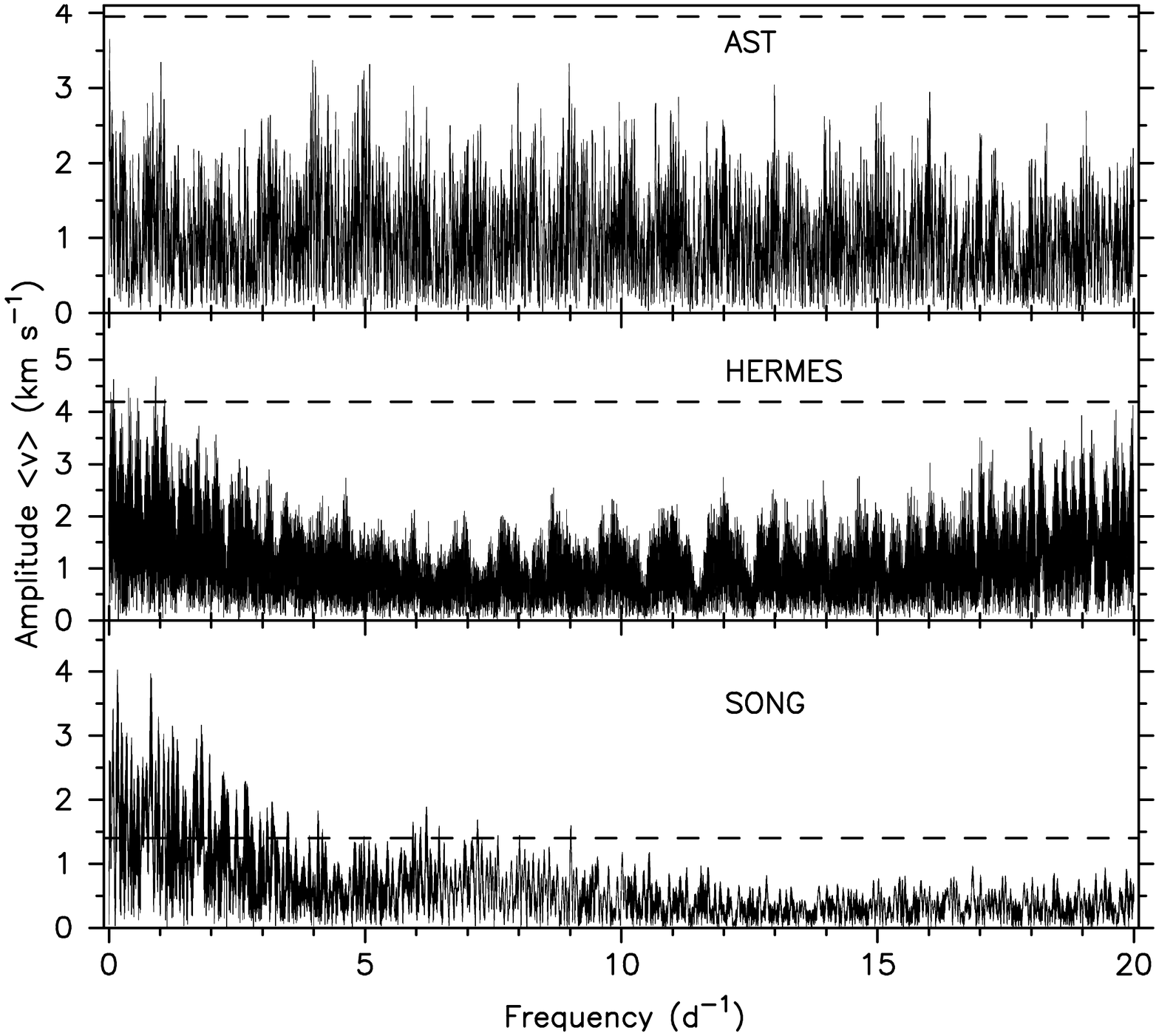}}}
%\vspace{0.3cm} 
\rotatebox{0}{\resizebox{11.cm}{!}{\includegraphics{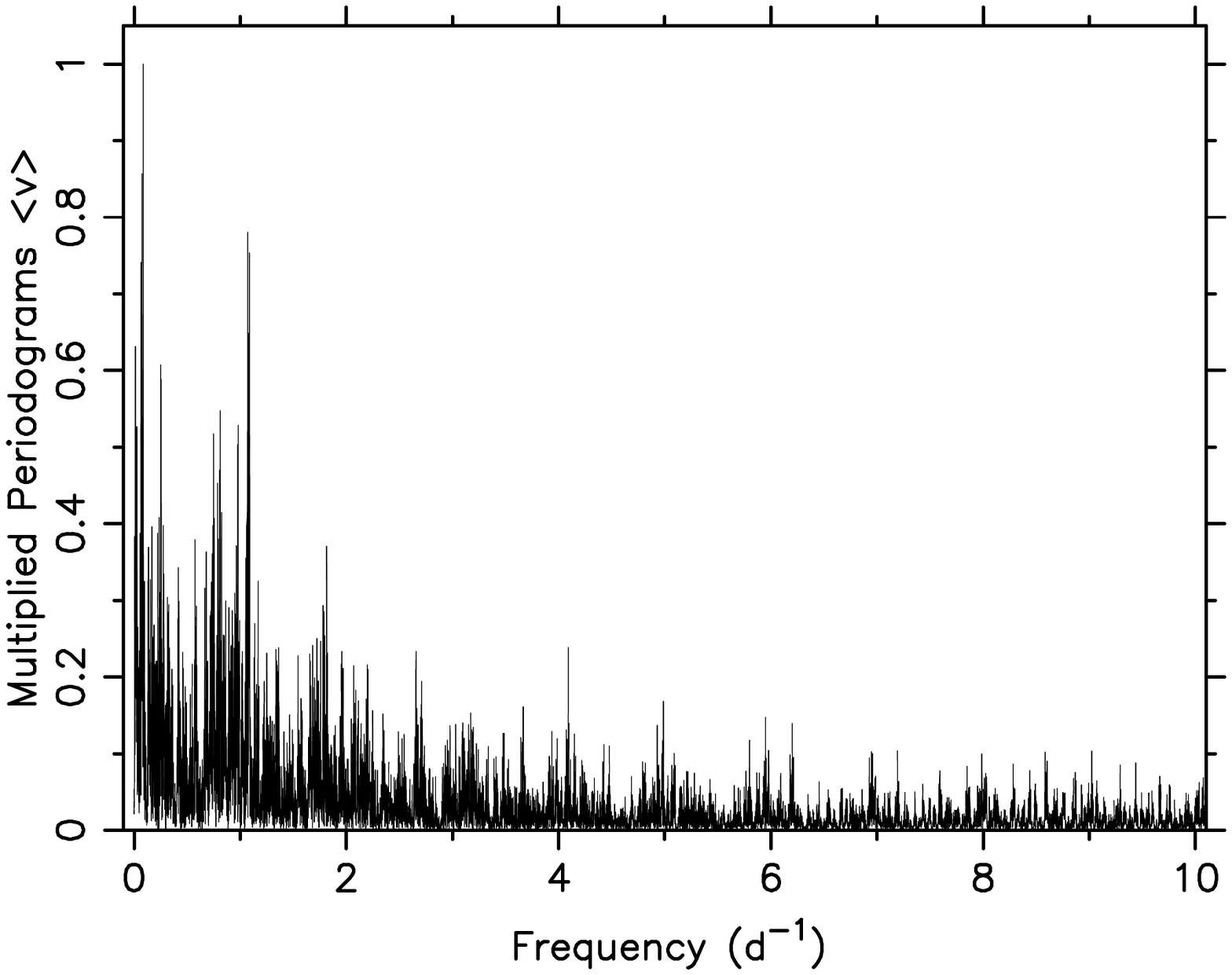}}}
\end{center}
\caption{Same as Fig.\,\ref{Scargle-ew-5875} but for $\langle v\rangle$ of the
  \ion{He}{i}\,5875\AA\ line.}
\label{Scargle-vrad-5875}
\end{figure}
\begin{figure}
\begin{center}
\rotatebox{270}{\resizebox{6.5cm}{!}{\includegraphics{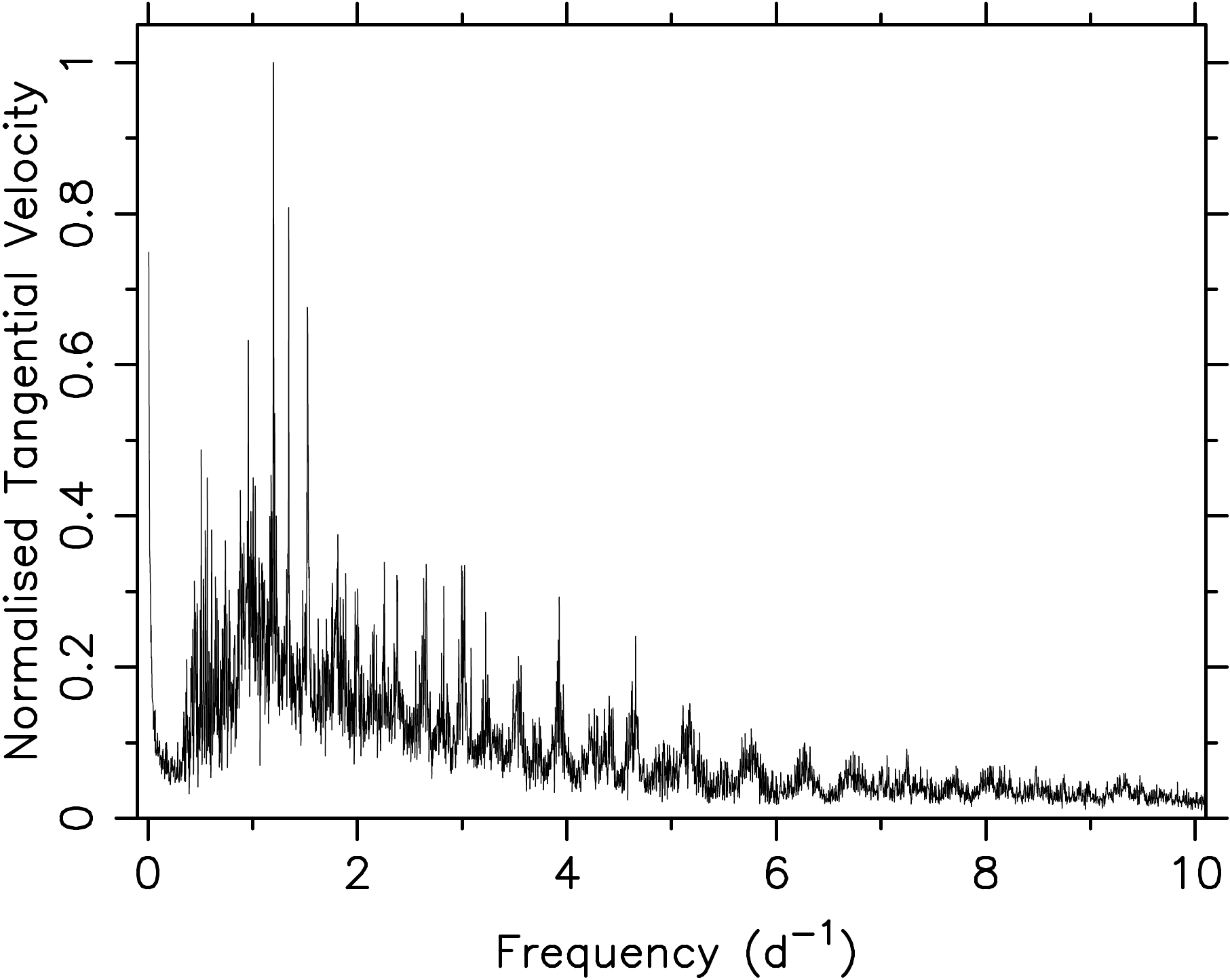}}}
\end{center}
\caption{Normalised periodogram for the tangential velocity of internal
  gravity waves occurring in 2D simulation D11 in \citet{Rogers2013}, properly
  scaled in frequency to take into account the mass and evolutionary status of
  HD\,188209 with respect to the simulated stellar model.}
\label{igw}
\end{figure}
\begin{figure}
\begin{center}
\rotatebox{0}{\resizebox{9.5cm}{!}{\includegraphics{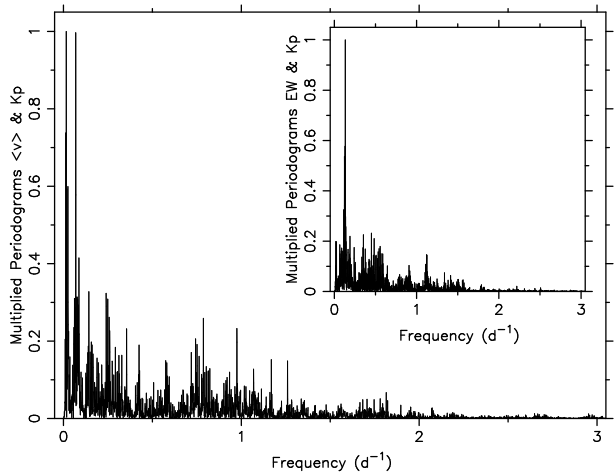}}}
\end{center}
\caption{Multiplied Scargle periodograms 
after weighting according to the SNR of
  the frequency with highest amplitude for the 
$\langle v\rangle$ (and EW in the inset) of the
  \ion{He}{i}\,5875\AA\ line and the {\it Kepler\/} photometry in Mask\,32.}
\label{Scargle-vrad-ew-Kp-5875}
\end{figure}

\begin{figure}
\begin{center}
\rotatebox{270}{\resizebox{6.5cm}{!}{\includegraphics{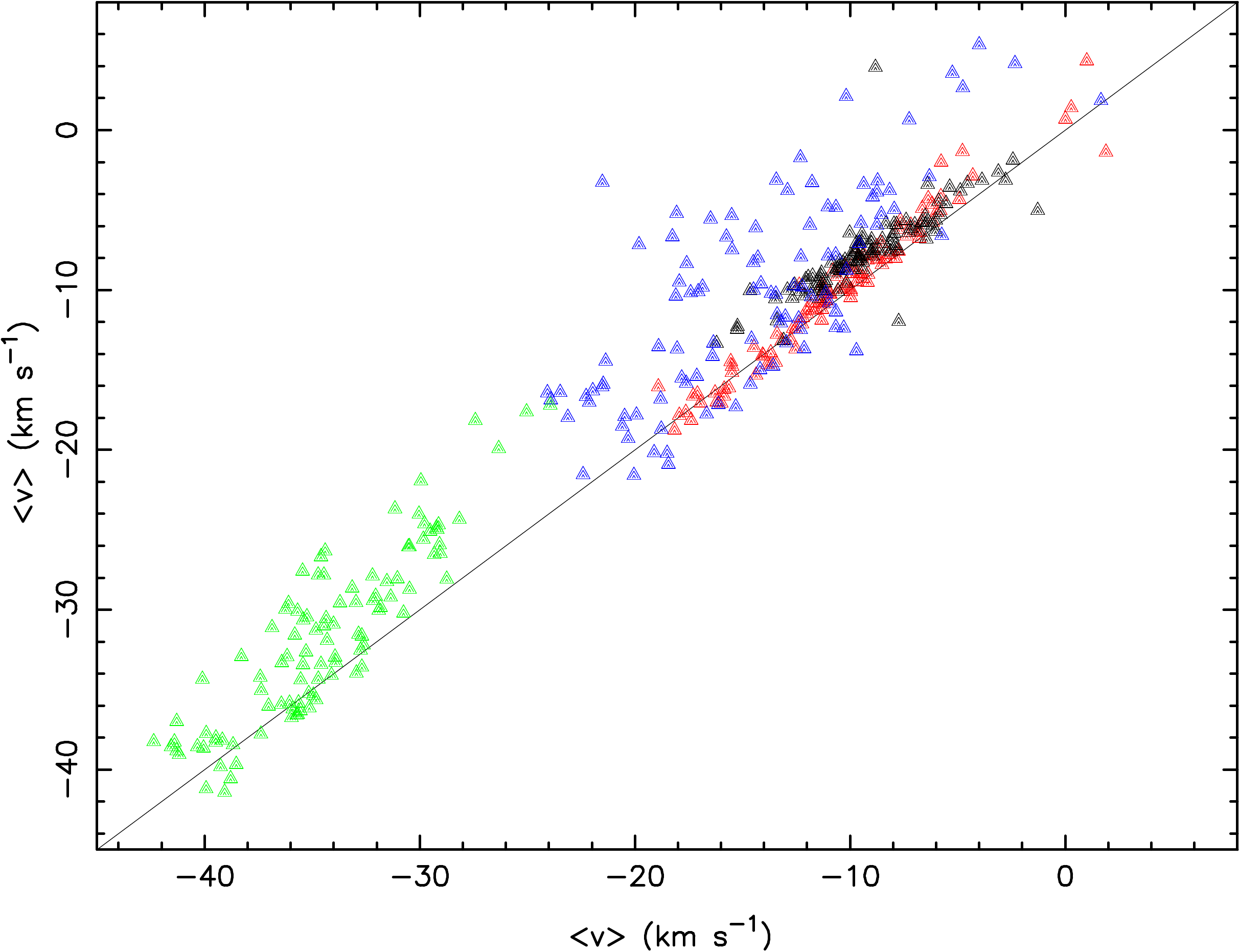}}}
\end{center}
\caption{The centroid velocities based on different spectral lines available in
  the HERMES spectroscopy: black: \ion{He}{ii}\,4541\AA\ versus
  \ion{He}{ii}\,5410\AA, red: \ion{He}{i}\,5015\AA\ versus \ion{He}{i}\,4922\AA,
  blue: \ion{He}{i}\,5875\AA\ versus \ion{Si}{iii}\,4567\AA, green: H$\beta$
  versus H$\gamma$.}
\label{biplot}
\end{figure}

\begin{figure*}
\begin{center}
\rotatebox{270}{\resizebox{12.cm}{!}{\includegraphics{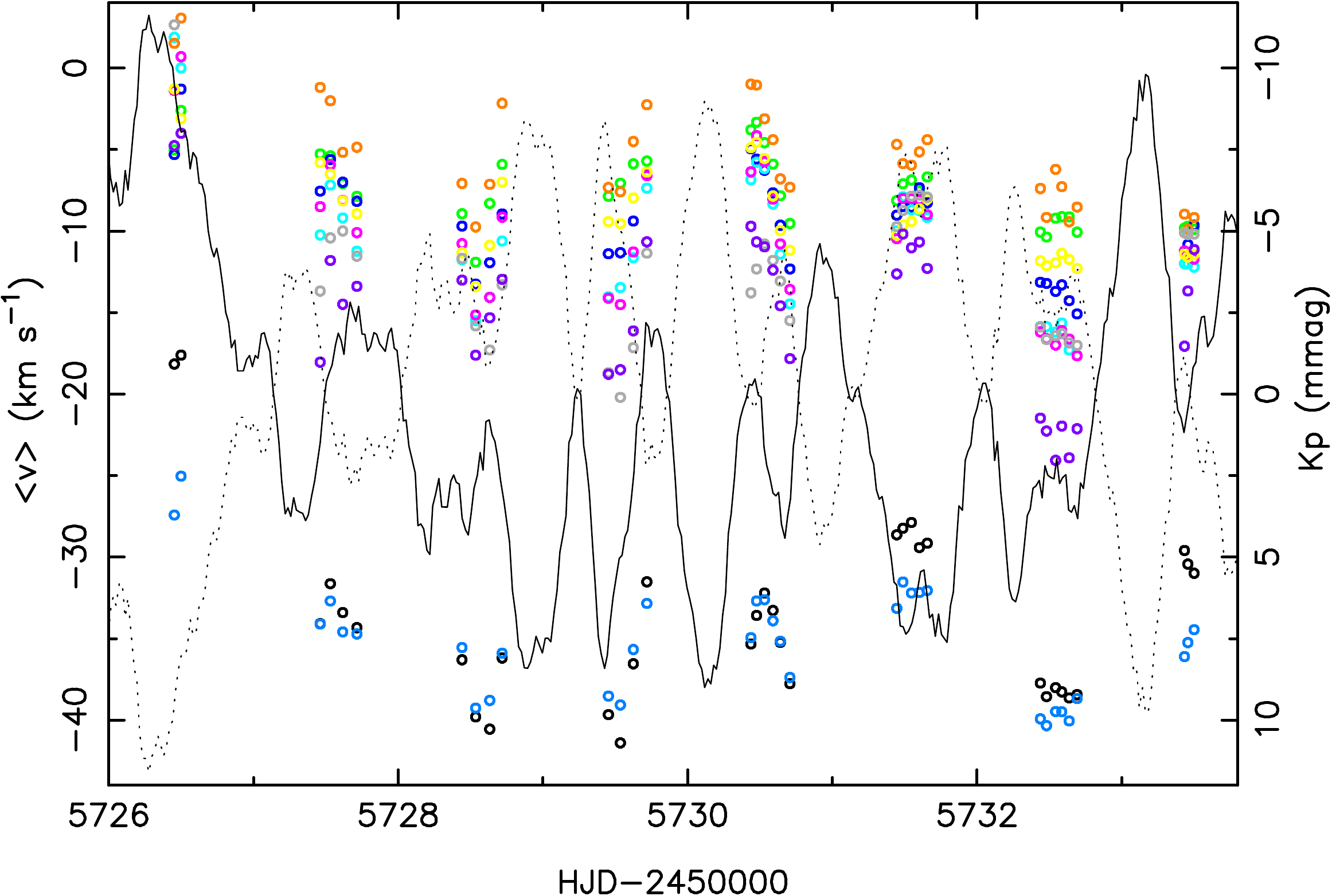}}}
\end{center}
\caption{The centroid velocity (left $y$-axis)
and {\it Kepler\/} photometry in Mask\,45 (right $y$-axis) during
  the only epoch where the space and ground-based data overlap in time. The
  colour coding is as follows: H$\beta$: black, H$\gamma$: light blue,
  \ion{He}{i}\,4713\AA: dark blue, \ion{He}{i}\,4922\AA: cyan,
  \ion{He}{i}\,5015\AA: pink, \ion{He}{i}\,5875\AA: grey,
  \ion{He}{ii}\,4541\AA: green, \ion{He}{ii}\,5410\AA: yellow,
  \ion{C}{iv}\,5812\AA: orange, \ion{Si}{iii}\,4567\AA: purple.  To
  guide the eye, the observed {\it Kepler\/} light curve is shown as dotted line while
  its mirrored version with respect to the average {\it Kepler\/} magnitude  
is shown as the full line.}
\label{vrad-Kp-all}
\end{figure*}
\begin{figure*}
\begin{center}
\rotatebox{270}{\resizebox{15.cm}{!}{\includegraphics{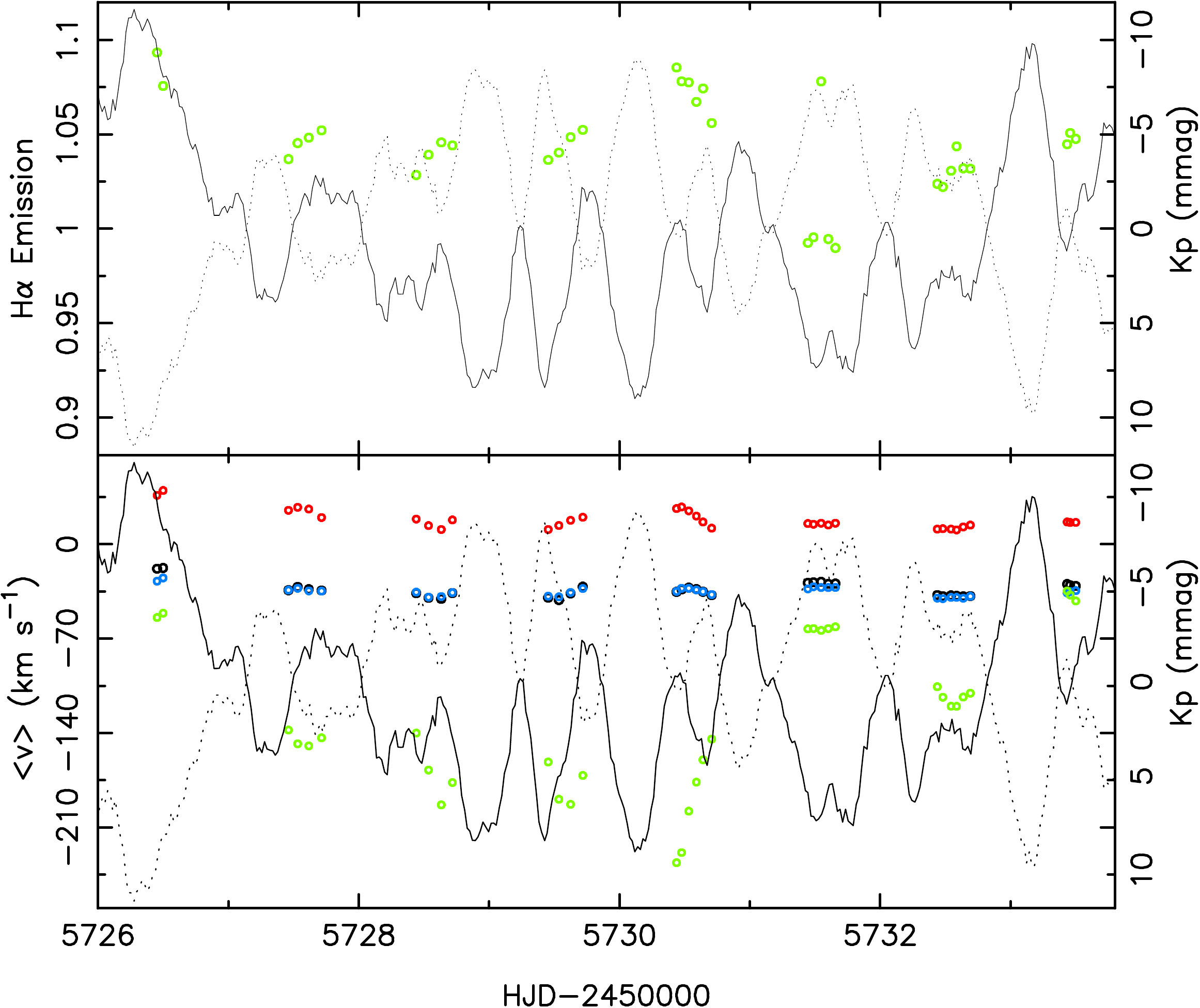}}}
\end{center}
\caption{The centroid velocity (left $y$-axis, bottom panel) and H$\alpha$
  emission strength expressed in continuum units (left $y$-axis, top panel), as
  well as the {\it Kepler\/} photometry in Mask\,45 (right $y$-axes) during the
  only epoch where the space and ground-based data overlap in time. The The
  colour coding is as follows: H$\alpha$: light green, H$\beta$: black,
  H$\gamma$: light blue, \ion{He}{i}\,4686\AA: red.  To guide the eye, the
  observed {\it Kepler\/} light curve is shown as dotted line while its mirrored
  version with respect to the average {\it Kepler\/} magnitude is shown as the
  full line.}
\label{vrad-Kp-wind}
\end{figure*}

These four instruments were constructed with different purposes and are attached
to telescopes of different sizes.  The specific
aim of HERMES is to obtain maximal capacity to detect time-resolved line-profile
variability such that the SNR is important. FIES is a similar spectrograph but
delivers lower resolution and is less well temperature stabilised.  The main
goal of SONG is to gain maximum precision in radial-velocity variations by means
of an iodine cell for solar-like oscillation studies \citep{Grundahl2017}, but
it also offers a ThAr option, which we have used here.  As we will show below
(see Figs\,\ref{Scargle-ew-5875} and \ref{Scargle-vrad-5875}) the lower
resolving power of AST, combined with the lower SNR it delivers, imply that these
data are at the limit of what we need to properly detect and interpret the
variability of HD\,188209, but adding this data set gave compatible results so
we kept them in the analysis and adopted an approach taking into account the
SNR.  Indeed, the SNR levels achieved with the four instruments are
different for similar integration times and provide us with a way to properly
weigh the data when intepreting the variability.

In the following, we focus mainly on those spectral lines that are available for
the four independent spectroscopic data sets listed in Table\,\ref{logbook} and
that do not suffer from order merging uncertainties in their line wings.  This
is the case for nine spectral lines.  We also considered the
\ion{Si}{iii}\,4567\AA\ line for its strong diagnostic power of stellar
oscillations \citep{Aerts2004} and of surface spots \citep{Briquet2004} of hot
stars, although it is not available in all data sets.  In addition, despite it
being available only in the HERMES data due to too poor order merging for the
other three instruments, we considered H$\gamma$ as important diagnostic of the
gravity behaviour of the star.

For each spectral line, we computed its line moments as good diagnostics for
variability by adopting their definition by \citet{Aerts1992}. It is crucial to
define the optimal integration limits in the line wings when computing the
moments for the detection and interpretation of low-amplitude line-profile
variability (e.g., Chapters 5 and 6 in \citet{Aerts2010} and \citet{Zima2008},
for extensive descriptions and publicly available software). Since we aim to
merge the moments of the same spectral lines obtained with four
different instruments, we determined the most optimal integration limits per
line by careful visual inspection after overplotting the entire time series.  In
this paper, we concentrate on the equivalent width of the lines (the moment of
order zero, hereafter abbreviated as EW) and on their heliocentric centroid
velocity (the first-order moment, denoted as $\langle v\rangle$). Typical
uncertainties for the centroid velocities due to uncertainties in the
integration limits and the noise in the spectral line range from 0.1 to
0.4\,km\,s$^{-1}$, while the systematic uncertainty due to limited knowledge of
the laboratory wavelength can reach up to a few km\,s$^{-1}$. The latter is not
of importance when studying {\it time variability\/} of one and the same
particular spectral line. The systematic uncertainty is only of importance when
comparing the average value of the centroid velocity for different spectral
lines to interpret their amplitude as a function of the formation depth in the
stellar photosphere or wind.

The atomic data are not equally good for different spectral lines. In view of
this, it is highly advantageous to make a line-by-line analysis per instrument
and combine the line diagnostic values in {\it Fourier domain\/} rather than in
the time domain. Indeed, merging in the time domain suffers from slight
inconsistencies in Doppler velocities derived from adopted laboratory
wavelengths for various lines and from imperfections in the normalisation of the
continuum flux, while merging in the Fourier domain is not affected by that.
Such an approach also allows to combine the temporal variability from entirely
different quantities derived from photometry and spectroscopy, with the
opportunity to reveal low-amplitude variations that appear consistently in
different types of independent data but with too low SNR in each of them
separately to be significant.

The method of combining data in the Fourier domain was already applied and
illustrated by \citet{Aerts2006} to discover previously unknown low-amplitude
non-radial oscillation modes in the archetype monoperiodic $\beta\,$Cep star
$\delta\,$Ceti from the combination of MOST space photometry and high-resolution
spectroscopy.  The rationale behind it is that the discrete Fourier transforms
of different time series data of the same variable star all achieve a maximum at
frequencies that are present in the data, while spurious frequencies due to
noise or aliasing are different for the various independent data sets. In this
respect, it is particularly useful to combine data sets with a different
sampling rate, as is the case for the SONG and HERMES data. Multiplication of
the Fourier transforms of the different data sets, after normalising them
according to the frequency of maximal amplitude, implies that true frequencies
get higher multiplied dimensionless amplitude while spurious frequencies tend to
cancel each other.  In the case of $\delta\,$Ceti mentioned above, it concerned
frequencies of standing waves due to coherent heat-driven oscillation
modes. However, the same principle also applies to isolated frequencies of
standing waves excited stochastically as in sun-like stars and red giants, or to
frequencies belonging to an entire spectrum of travelling waves. Whatever their
excitation mechanism, standing waves or travelling waves inside a star have
frequency values connected with the stellar structure properties and so they
occur at the same values in the various data sets.

\subsection{The showcase of the \ion{He}{i}\,5875\,\AA\  line}

We first considered the \ion{He}{i}\,5875\AA\ to illustrate the variability
behaviour as this line leads to the highest significance for the variability in
terms of SNR in the line diagnostics in the four spectroscopic data sets.
The individual zeroth and first moment variations of the
\ion{He}{i}\,5875\AA\ line are shown in Fig.\,\ref{vrad-ew-5875}. It can
be seen that slight differences occur in the EW of the line for the four
spectrographs but also for the same spectrograph during different epochs. This
may in part be due to imperfect normalisation, but it may also be caused by the
rather irregular variability of the star, as revealed in the {\it Kepler\/} data in
Fig.\,\ref{LCs}.  Even though the moments in the definition by \citet{Aerts1992}
are specifically defined such as to compensate optimally for small differences
in the EW (cf.\ lower panel of Fig.\,\ref{vrad-ew-5875}), it is 
advantageous to treat the four $\langle v\rangle$ data sets separately, given
that the four average values of this quantity per spectrograph is not equal and
merging in the time domain prior to the computation of the Fourier transform
would introduce spurious low frequencies.

The individual Scargle periodograms \citep{Scargle1982} for the HERMES, SONG,
and AST data sets of the EW and $\langle v\rangle$ of the \ion{He}{i}\,5875\AA\ line are
shown in the upper three panels of Figs\,\ref{Scargle-ew-5875} and
\ref{Scargle-vrad-5875} (the FIES data set being too scarce for this
purpose). As anticipated, these periodograms suffer from daily
aliasing, a well-known phenomenon in time-series analysis of single-site data
sets.  The $\langle v\rangle$ based on the AST data do not 
lead to significant
frequencies (maximum amplitude typically between 3 and 4 times the SNR), but the
HERMES and SONG data sets do reveal significant low-frequency amplitude, with a
factor typically between 10 to 12 times the SNR for SONG and 4 to 6 times the
SNR for HERMES (Fig.\,\ref{Scargle-vrad-5875}).  For the EW, the significance
levels are typically somewhat lower than for the $\langle v\rangle$
(Fig.\,\ref{Scargle-ew-5875}).

For each of these individual periodograms, the maximum amplitude was sought and
transformed to its value expressed as a function of the average SNR, where the
latter was computed in the Scargle periodogram over the frequency range
$[10,20]\,$d$^{-1}$. The dashed lines in Figs\,\ref{Scargle-ew-5875} and
\ref{Scargle-vrad-5875} are placed at four times this SNR Subsequently, the
three Scargle periodograms expressed in units of the SNR were multiplied, after
which they were normalised such that the maximum peak after this multiplication
was placed at value 1.0. The results of this procedure are shown in the lower
panels of Figs\,\ref{Scargle-ew-5875} and \ref{Scargle-vrad-5875}. It can be
seen that this procedure leads to a far clearer view of the variability in this
single spectral line, the aliasing and noise peaks in the individual three data
sets being substantially reduced in this way.  Variability is found to occur at a
multitude of low frequencies, in the range $[0,2]\,$\,d$^{-1}$, entirely in
agreement with the {\it Kepler\/} scattered-light photometry.

The morphology of the normalised periodogram in Fig.\,\ref{Scargle-vrad-5875} is
unlike those encountered for B dwarfs pulsating multiperiodically in coherent
heat-driven gravity modes \citep[e.g.,][]{DeCatAerts2002}. In view of this
discrepancy and the similar amplitude spectrum obtained from the {\it Kepler\/}
photometry, we compare the frequency spectrum of HD\,188209 in
Fig.\,\ref{Scargle-vrad-5875} with predictions from hydrodynamical simulations
based on convectively-driven internal gravity waves. In order to do so, we
follow a similar approach as in \citet{AertsRogers2015}; only here we have a
simpler case because the 2D simulations by \citet{Rogers2013} provide us with
velocity information that allows a direct meaningful comparison with
Fig.\,\ref{Scargle-vrad-5875}, while \citet{AertsRogers2015} had to transform
this information to brightness variations to be able to compare with CoRoT light
curves. Given that HD\,188209 is evolved and about ten times more massive than
the model of 3\,M$_\odot$ adopted for the simulations by \citet{Rogers2013}, we
performed a scaling with a factor 0.53 that occurs between the frequencies of
dipole gravity waves of a 3\,M$_\odot$ ZAMS star and of a 30\.M$_\odot$ TAMS
star \citep[][Table\,1]{Shiode2013}. Moreover, we normalise the tangential
velocity of the 2D simulations (the radial component being entirely negligible -
see Fig.\,1 in \citet{AertsRogers2015}, right panel), to their highest
amplitude, considering simulation run D11 (a non-rigidly rotating star whose
core rotates 1.5 times faster than its envelope). The outcome is presented in
Fig.\,\ref{igw}. While this exercise does not allow a peak-to-peak comparison of
the frequencies in the two spectra, the overall morphology in
Figs\,\ref{Scargle-vrad-5875} and \ref{igw} is similar.

Next, we went one step further and considered a multiplied Scargle periodogram
deduced from the three spectroscopic \ion{He}{i}\,5875\AA\ data sets and the
{\it Kepler\/} data in Mask\,32 as a fourth independent data set. Its highest
amplitude peak has a significance of 137 times the SNR in the frequency range
$[10,20]\,$d$^{-1}$, the average SNR being 27\,ppm in that interval.  The
outcome of this multiplication is shown in
Fig.\,\ref{Scargle-vrad-ew-Kp-5875}. Figures\,\ref{Scargle-ew-5875},
\ref{Scargle-vrad-5875}, and \ref{Scargle-vrad-ew-Kp-5875} all point towards the
same conclusion: HD\,188209 reveals significant variability with an entire
spectrum of significant frequencies below 2\,d$^{-1}$. The morphology in the
periodograms point towards travelling waves at the origin of this variability,
given the density of frequency peaks and the absence of any clear relationship
between the dominant peaks.  Thus, we find fully compatible variability results
between the {\it Kepler\/} scattered-light space photometry and the
\ion{He}{i}\,5875\AA\ spectral line, both when treating these data separately
and in a combined analysis.

\subsection{Other spectral lines}

We repeated the entire procedure described in the previous section for the other
photospheric lines that are present in all four spectroscopic data sets. It
concerns \ion{He}{i}\,4713\AA, \ion{He}{i}\,4922\AA, \ion{He}{i}\,5015\AA, \ion{He}{ii}\,4541\AA, \ion{He}{ii}\,5410\AA,
\ion{C}{iv}\,5812\AA.  Moreover, we considered the \ion{Si}{iii}\,4567\AA\ line in the HERMES
spectra.  {\it All these lines behave fully consistently\/} with the
\ion{He}{i}\,5875\AA\ line in terms of time-variability patterns and morphology of the
separate and multiplied Fourier transforms, the only difference being the
amplitudes of their EW and $\langle v\rangle$ time series (not the relative
amplitudes expressed in terms of the SNR, which are remarkable similar for all
those lines). Figure\,\ref{biplot} shows a comparison of the
$\langle v \rangle$-values for several lines available in the HERMES
spectroscopy, where we plot them line-by-line. It can be seen that there is
large consistency among the values of the centroid velocities.  The values of
$\langle v\rangle$ for the \ion{He}{i} lines have the strongest similarity. The
\ion{Si}{iii}\,4567\AA\ line is more affected by the base of the wind than the
\ion{He}{i}\,5875\AA\ line and the Balmer lines are clearly formed furher out in the
wind than the helium and metal lines.

Next, we considered two lines available in all four data sets but formed
partially in the wind and in the photosphere: H$\beta$ and
\ion{He}{ii}\,4686\AA.  The latter spectral line shows an unexplained
global shift to the red with an average velocity value of about
+45\,km\,s$^{-1}$ with respect to all other spectral lines when considering its
laboratory wavelength of 4685.71\AA.  This anomaly occurs for all four the
spectroscopic data sets.  It is known that the \ion{He}{ii}\,4686\AA\ line
is partially formed in the stellar wind
\citep[e.g.,][]{Massey2004,Martins2015a}. As such, its formation is affected by
the temperature, wind density, metallicity, and amount of EUV flux. However, we
cannot explain the detected redward shift corresponding with some 0.7\AA\ that
we find for the absorption profile of this line. We are aware of at least one
other star where this line is in absorption and for which the same anomaly is
reported in the literature (see Fig.\,15 in \citet{Massey2004} for the O8.5 I(f)
supergiant AV\,469 in the Small Magellanic Cloud, a star with quite similar
fundamental parameters than HD\,188209 except for the metallicity). Recent
laboratory measurements of \ion{He}{ii} pointed out that its spectral line structure
near 4686\AA\ is complex \citep{Syed2012}. Caution of its interpretation for
absorption lines of hot evolved stars seems in order.

The H$\beta$ line of HD\,188209 behaves very similarly to the H$\gamma$ line,
the latter only being suitable for our analysis in the HERMES data set. This is also the
case for the H$\alpha$ and \ion{Si}{iii}\,4567\AA\ lines, which we only considered 
in the HERMES data set. Apart from the global shift of its average centroid
velocity with respect to the other lines, the variability behaviour of the
\ion{He}{ii}\,4686\AA\ line is fully consistent with that of the other spectral lines
and confirms that this line is partially formed in the wind.

Unfortunately, there is only one short period spanning eight days for which we
have both {\it Kepler\/} space photometry and ground-based spectroscopy. It
concerns data in Mask\,45 and HERMES spectroscopy. The centroid velocities of
all the available spectral lines and the space photometry taken during these
eight days are shown in Fig.\,\ref{vrad-Kp-all}. The \ion{He}{ii}\,4686\AA\
line and H$\alpha$ fall outside the boundaries of this plot. It is seen that 
{\it all spectral lines show consistent behaviour in their centroid velocity in
  the time domain}. 

The \ion{He}{ii}\,4686\AA\ line is shown along with H$\alpha$, H$\beta$,
and H$\gamma$, and in comparison with the {\it Kepler\/} photometry, in
Fig.\,\ref{vrad-Kp-wind}. It can be seen that \ion{He}{ii}\,4686\AA,
H$\beta$, and H$\gamma$ are fully in agreement with each other in terms of
variability. The H$\alpha$ emission varies in antiphase with the H$\alpha$
centroid velocity. While the relation between the spectroscopic variability and
the {\it Kepler\/} photometry is hard to unravel from Figs\,\ref{vrad-Kp-all}
and \ref{vrad-Kp-wind}, the morphology of the various quantities in the time
domain is similar. {\it The spectroscopic variability of the wind and of
  the photosphere and the photometric variability are all in agreement in terms
  of amplitudes and periodicities. }

\subsection{Summary of variability of the centroid velocities}

\begin{figure*}
\begin{center}
\rotatebox{270}{\resizebox{11cm}{!}{\includegraphics{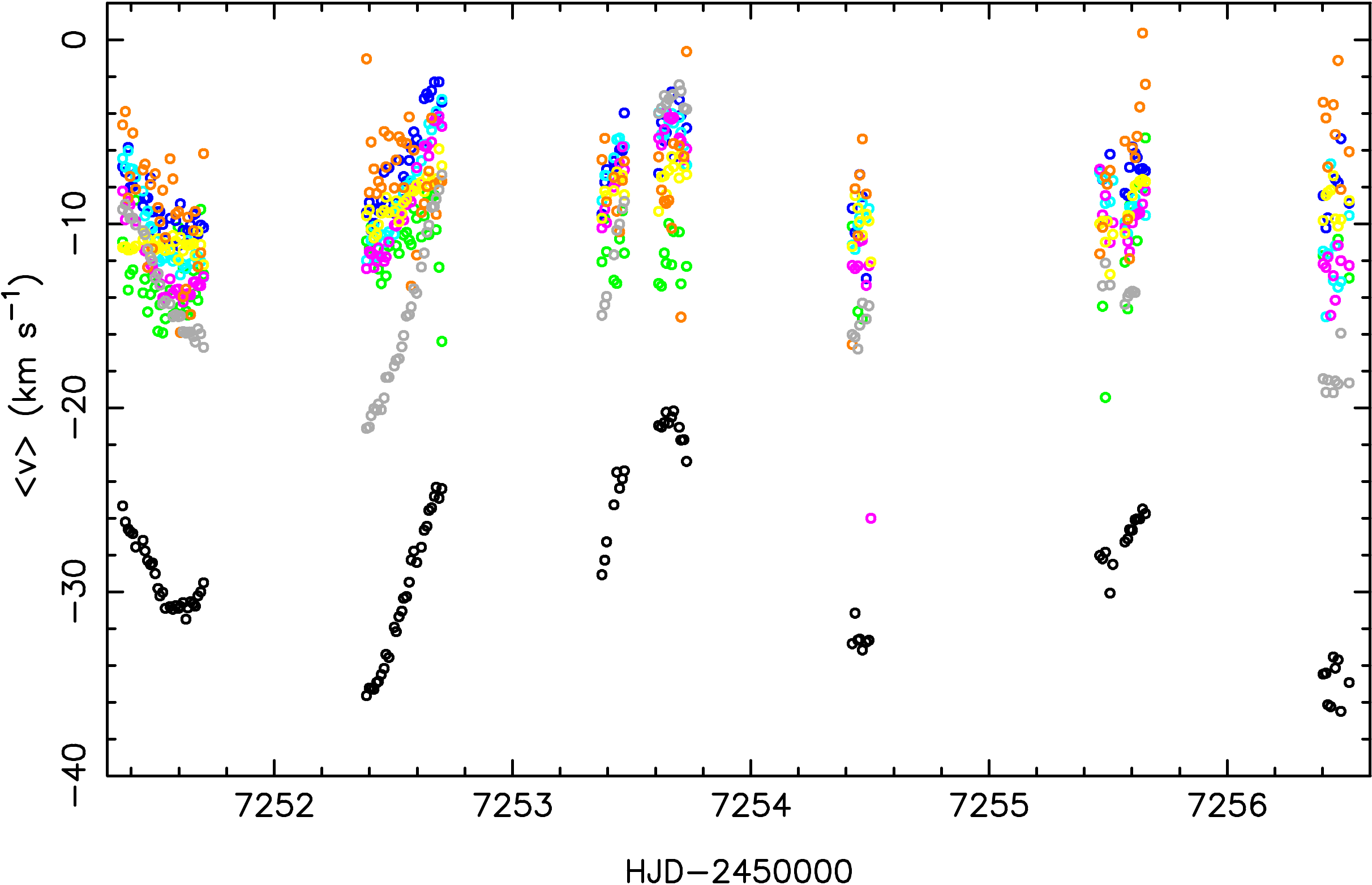}}}
\end{center}
\caption{Excerpt of the SONG spectroscopy during 5 consecutive days. The centroid
  velocity is shown for nine available spectral lines (the 
\ion{He}{ii}\,4686\AA\ line has been avoided for visibility purposes),
The colour coding is the
  same as in Fig.\,\protect\ref{vrad-Kp-all}.}
\label{vrad-song-all}
\end{figure*}

Among the spectroscopic data sets, the highest sampling rate is obtained with
SONG. A 5\,d excerpt of $\langle v\rangle$ derived from nine spectral lines
available in these data is shown in Fig.\,\ref{vrad-song-all}. We find similar
behaviour for most of the spectral lines. The figure reveals different behaviour
in variability from night to night, with dominant periodicities typically longer
than 10\,h and quite different amplitudes per night. This confirms the earlier
finding of variability with frequencies below 2\,d$^{-1}$ in the Fourier domain,
but now by visual inspection in the time domain based on the spectral lines
available in the SONG spectroscopy.
\begin{figure*}
\begin{center}
\rotatebox{270}{\resizebox{11cm}{!}{\includegraphics{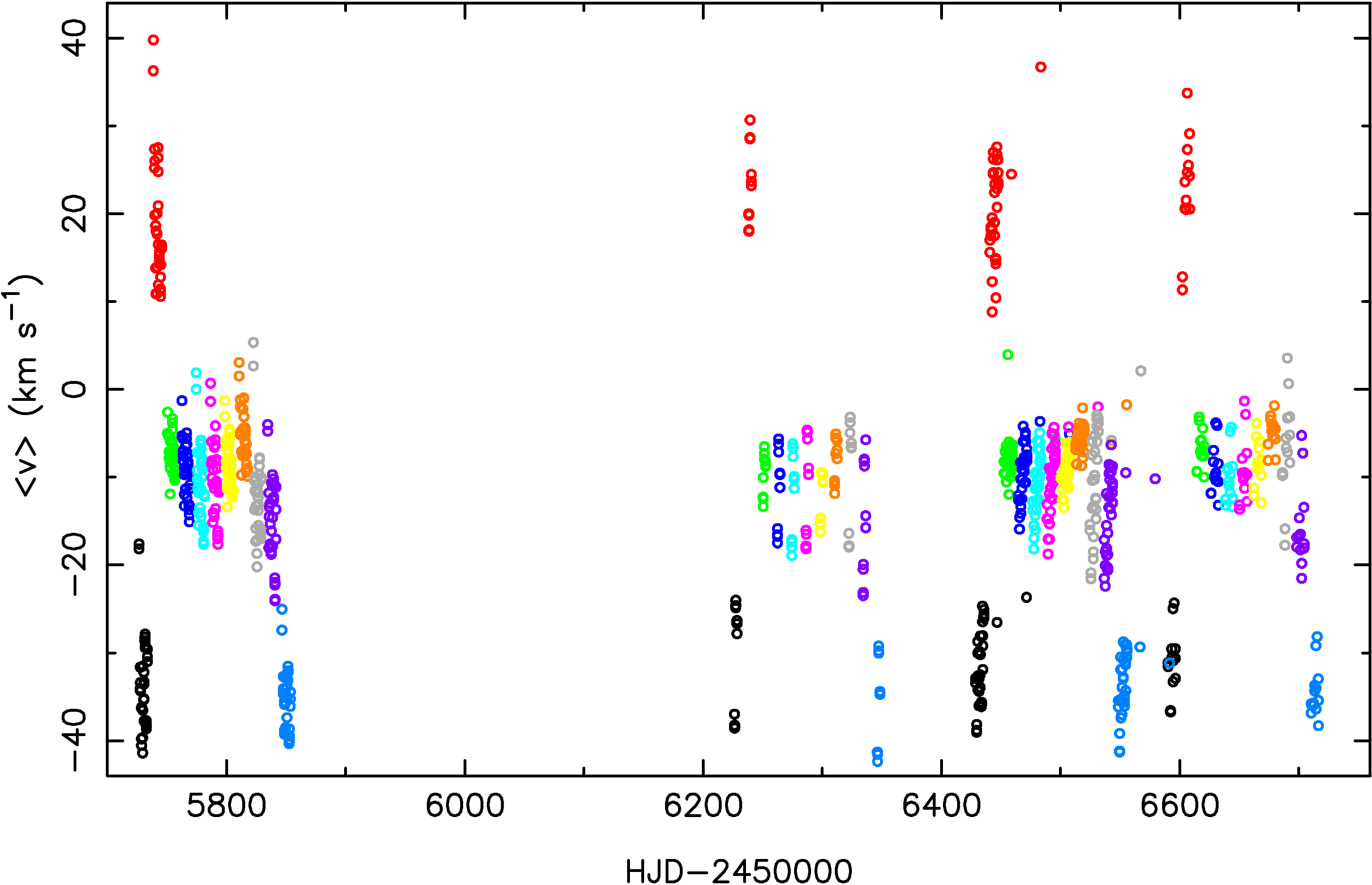}}}
\end{center}
\caption{HERMES spectroscopy of HD\,188209 covering some 870 days of
  monitoring during four epochs. The centroid velocity is shown for all
  available spectral lines except H$\alpha$.  The colour coding is the same as
  in Fig.\,\protect\ref{vrad-Kp-all}. For visibility purposes, the values for
  each spectral line have been shifted in time with multiples of 12\,d, H$\beta$ in
  black representing the true HJD.}
\label{vrad-hermes-all}
\end{figure*}

Thanks to its construction, the HERMES spectrograph offers the best long-term
stability of the four instruments used here. Indeed, its fiber-fed design
connected with a temperature-controlled room and its dedicated calibration
pipeline were specifically defined and implemented with long-term monitoring of
variable phenomena in mind \citep{Raskin2011}.  This data set of HD\,188209 is
therefore best suited to illustrate the long-term spectroscopic variability of
the star in the time domain. We show the centroid velocities for the eleven
available spectral lines in the HERMES data for the first four epochs of
monitoring in Fig.\,\ref{vrad-hermes-all}, where the HJD for the various
spectral lines have been shifted along the $x-$axis of the plot with multiples
of 12 days for visibility purposes.  The data for the first epoch in
Fig.\,\ref{vrad-hermes-all} corresponds with that shown in
Fig.\,\ref{vrad-Kp-all}. Figures\,\ref{vrad-Kp-all} and \ref{vrad-hermes-all}
illustrate the consistency between the $\langle v\rangle$-values for the various
spectral lines, both on a time scale of a week and of two years.

\begin{table*}[h!]
  \caption{Summary of variability properties of the centroid velocity 
    $\langle v\rangle$ and EW    for twelve 
    spectral lines of HD\,188209 available in the HERMES data, spanning 1800\,d.}
\label{alllines}
\centering 
\tabcolsep=2pt                                    
\begin{tabular}{clcccccccc}          
\hline\\[-10pt]     
Spectral & Laboratory & Colour &
Minimum & Maximum & Range & Average &  
Minimum & Maximum & Range \\
Line & Wavelength (\AA)\tablefootmark{1} & Code & $\langle v\rangle$ (km\,s$^{-1}$) 
& $\langle v\rangle$ (km\,s$^{-1}$)
& $\langle v\rangle$ (km\,s$^{-1}$) & $\langle v\rangle$ (km\,s$^{-1}$) & EW (\AA) & EW (\AA) & 
EW (\AA) \\
\hline
H$\alpha$ & 6562.801 & light green & -292.4 & +439.6 & 732.0 & -85.6 & -0.163 & 0.694 & 0.857 \\ 
H$\gamma$ & 4340.464 & light blue & -42.4 & -23.9 & 18.5 & -34.4 & 1.353 & 1.553 & 0.201 \\
H$\beta$ & 4861.325&  black & -41.4 & -17.2 & 24.2 & -31.7 & 1.249 & 1.631 & 0.382 \\
\ion{Si}{iii} & 4567.841 & purple &-24.1 & +1.7 & 25.8 & -14.4 & 0.094 & 0.137 & 0.044 \\
\ion{He}{i} & 4922.931 & cyan & -18.9 & +1.9 & 20.8 & -10.6 & 0.495 & 0.589 & 0.094 \\
\ion{He}{i} & 5875.599 & grey & -21.6 & +5.4 & 27.1 & -10.2 & 0.987 &  1.378 & 0.391 \\
\ion{He}{i} & 5015.678 & pink & -18.8 & +4.3 & 23.1 & -10.0 & 0.328 & 0.391 & 0.063 \\
\ion{He}{i} & 4713.139 & dark blue & -17.5 & +0.2 & 17.7 & -9.4 & 0.325 & 0.410 & 0.085 \\
\ion{He}{ii} & 5410.53 & yellow & -16.2 & -1.3 & 14.9 & -9.2 & 0.363 & 0.398 & 0.035 \\ 
\ion{He}{ii} & 4541.591 & green & -13.3 & +3.9 & 17.3 & -7.6 & 0.220 & 0.311 & 0.091 \\
\ion{C}{iv} & 5811.97 & orange & -11.9 & +4.4 & 16.3 & -5.5 & 0.085 & 0.107 & 0.022 \\
\ion{He}{ii} & 4685.71 & red & +8.8 & +43.7 & 34.8 & 20.8 & 0.187 & 0.253 &0.066\\
\hline
\end{tabular}
\tablefoot{
  \tablefoottext{1}{
    Retrieved from the Atomic Line List available at {\tt
      http://www.pa.uky.edu/$\sim$peter/newpage/.}} Systematic uncertainties in 
  the average values of 
  $\langle v\rangle$ are different for the different spectral lines and may
  amount to several km\,s$^{-1}$ per spectral line. This is typically an order
  of magnitude larger than the
  statistical uncertainties for each of the $\langle v\rangle$-values for a fixed
  spectral line.}
\end{table*}

A summary of all the computed line diagnostic is provided in
Table\,\ref{alllines} for the HERMES data for which the centroid velocities have
been plotted in Fig.\,\ref{vrad-hermes-all} (except for H$\alpha$ as these do
not fit the plot).  The colour coding adopted in Figs\,\ref{vrad-Kp-all},
\ref{vrad-Kp-wind}, \ref{vrad-song-all}, and \ref{vrad-hermes-all} has been
indicated and the lines are listed in order of increasing average centroid
velocity, corresponding to increasing line-formation depth into the
wind/photosphere. Despite unknown systematic uncertainties for
$\langle v\rangle$ due to limitations in the knowledge of the laboratory
wavelengths (of order km\,s$^{-1}$), we detect a radial gradient in the average
centroid velocities for the different lines (Table\,\ref{alllines}), where, as
already mentioned, the \ion{He}{ii}\,4686\AA\ line behaves anomalously (as also
shown in Fig.\,\ref{vrad-hermes-all}). The peak-to-peak variability of
$\langle v\rangle$ is very similar if one keeps in mind that line blending
affects this range and is different for the various spectral lines. Even though
they are formed partly in the wind in view of their more negative average
velocity, the variability of H$\beta$ and H$\gamma$ behaves similarly to the one
of the helium and metal lines.  Hence, the photospheric variability does not
change at the bottom of the stellar wind.  Only the H$\alpha$ variability seems
dominated by the wind behaviour.

%%%%%%%%%%%%%%%%

\section{Discussion and conclusions}

In this work, we have used the {\it Kepler\/} spacecraft far beyond its nominal
performance by studying scattered-light photometry of the bright blue supergiant
HD\,188209 while it was situated in between active CCDs. Aperture photometry of
its scattered light delivered the first four-years long uninterrupted
high-cadence light curve of a blue supergiant, reaching a precision of some
27\,ppm at frequencies above 10\,d$^{-1}$. We found similarities in the morphology
of the frequency spectrum derived from the scattered light and from line-profile
diagnostics of several spectral lines in long-term ground-based high-resolution
spectroscopy. A major conclusion of this work is that the range of detected
frequencies in the space photometry and in the ground-based spectroscopy is the
same.  All the frequency spectra point towards variability 
occurring in the photosphere and consistently propagating into the bottom
of the stellar wind.  The nature of the short-time Fourier transforms of the
high-cadence photometry excludes an interpretation in terms of standing waves
connected with non-radial gravity-mode oscillations, but rather points towards
the excitation of an entire spectrum of travelling internal gravity waves triggered
by core and/or envelope convection. This is a plausible explanation in terms of the
measured velocity amplitudes and the multitude of detected frequencies in the
regime below 2\,d$^{-1}$.

While we could not find a simple point-to-point relationship between the
photometric and spectroscopic data that were taken simultaneously, we did find
full consistency in the frequency range caused by these independent data
sets. Moreover, the overall morphology of the variable patterns in the time
domain and in the frequency spectra in Fourier space derived from the
scattered-light photometry and from the centroid velocities deduced from the
spectroscopy is very similar. The observed frequency spectra of HD\,188209 are
in qualitative agreement with those for the tangential velocities based on 2D
hydrodynamical simulations of internal gravity waves in a massive star. We thus
conclude to have found the first observational evidence of the occurrence of such
waves in a massive blue supergiant. Along with the discovery of such waves in
young O-type dwarfs \citep{Blomme2011,AertsRogers2015}, we revealed  at least one
important mechanism of angular momentum transport active in massive stars during
and beyond core-hydrogen burning that is currently not included in stellar
evolution models. It remains to be studied how much impact the omission of this
ingredient has for stellar evolution theory.

The tangential velocities associated with internal gravity waves in the stellar
photosphere of massive stars are of order a tenth of a km\,s$^{-1}$ per
individual wave \citep{Rogers2013}. It has already been shown by \citet[][their
Fig.\,5]{AertsRogers2015} that the collective effect of hundreds of such
internal gravity waves on line profiles is very similar to the effect due to a
collection of coherent heat-driven gravity-mode oscillations, particularly for
the line wings.  Given that both coherent standing gravity-mode oscillations and
running gravity waves have completely dominant tangential velocities (at the
level that their radial velocity component is negligible), any proper modelling
of macroturbulent line broadening due to gravity waves as detected in HD\,188209
requires fitting a tangential macroturbulent velocity field to the spectral
lines.

%________________________________________________________________

\begin{acknowledgements}
  This project has received funding from the European Research Council (ERC)
  under the European Union's Horizon 2020 research and innovation programme
  (Advanced Grant agreements N$^\circ$670519: MAMSIE ``Mixing and Angular
  Momentum tranSport in MassIvE stars'' and N$^\circ$267864: ASTERISK
  ``ASTERoseismic Investigations with SONG and Kepler'').  Funding for the
  Stellar Astrophysics Centre is provided by The Danish National Research
  Foundation (Grant DNRF106).  PIP acknowledges support from The Research
  Foundation Flanders (FWO), Belgium and EM was supported by the People
  Programme (Marie Curie Actions) of the European Union's Seventh Framework
  Programme FP7/2007-2013/ under REA grant agreement N$^\circ$623303 (ASAMBA).
  The authors are grateful to P.\ Beck, G.\ Holgado, S.\ Rodriguez, V.\ Schmid,
  and C.\ Gonzalez for having taken a few additional spectra included in this
  study.  Funding for the {\it Kepler\/} Discovery mission was provided by
  NASA’s Science Mission Directorate. The authors gratefully acknowledge the
  entire {\it Kepler\/} team, whose outstanding efforts have made these results
  possible.  This research has made use of the SIMBAD database, operated at CDS,
  Strasbourg, France, and of the Multimission Archive at STScI (MAST), USA.  We
  also acknowledge use of the Atomic Line List offered by the University of
  Kentucky, USA and maintained by Peter van Hoof, Royal Observatory of Belgium.
\end{acknowledgements}

%________________________________________________________________

\bibliographystyle{aa}	    % (uses file "plain.bst")
\bibliography{hd188209}		% expects file "myrefs.bib"

\end{document}